\documentclass[10pt]{iopart}
\usepackage{cite}
\usepackage{iopams}  
\usepackage{braket}
\usepackage{wasysym}
\usepackage[hidelinks]{hyperref}
\usepackage{subcaption}
\usepackage{tikz-cd}
\usepackage[shortlabels]{enumitem}
\usepackage[left=1.5cm,right=1.5cm,top=2cm,bottom=1.5cm]{geometry}

\graphicspath{{./}}

\begin{document}

\title[Self-Assembly of Geometric Space from Random Graphs]{Self-Assembly of Geometric Space from Random Graphs}

\author{Christy Kelly$^1$, Carlo A. Trugenberger$^2$ \& Fabio Biancalana$^1$}

\address{$^1$ School of Engineering and Physical Sciences, Heriot-Watt University, Edinburgh EH14 4AS, UK}
\address{$^2$ SwissScientific Technologies SA, rue du Rhone 59, CH-1204 Geneva, Switzerland}
\eads{\mailto{ckk1@hw.ac.uk}, \mailto{ca.trugenberger@bluewin.ch}, \mailto{f.biancalana@hw.ac.uk}}
\vspace{10pt}
\begin{indented}
\item[]April 2019
\end{indented}

\begin{abstract}
We present a Euclidean quantum gravity model in which random graphs dynamically self-assemble into discrete manifold structures. Concretely, we consider a statistical model driven by a discretisation of the Euclidean Einstein-Hilbert action; contrary to previous approaches based on simplicial complexes and Regge calculus our discretisation is based on the Ollivier curvature, a coarse analogue of the manifold Ricci curvature defined for generic graphs. The Ollivier curvature is generally difficult to evaluate due to its definition in terms of optimal transport theory, but we present a new exact expression for the Ollivier curvature in a wide class of relevant graphs purely in terms of the numbers of short cycles at an edge. This result should be of independent intrinsic interest to network theorists. Action minimising configurations prove to be cubic complexes up to defects; there are indications that such defects are dynamically suppressed in the macroscopic limit. Closer examination of a defect free model shows that certain classical configurations have a geometric interpretation and discretely approximate vacuum solutions to the Euclidean Einstein-Hilbert action. Working in a configuration space where the geometric configurations are stable vacua of the theory, we obtain direct numerical evidence for the existence of a continuous phase transition; this makes the model a UV completion of Euclidean Einstein gravity. Notably, this phase transition implies an area-law for the entropy of emerging geometric space. Certain vacua of the theory can be interpreted as baby universes; we find that these configurations appear as stable vacua in a mean field approximation of our model, but are excluded dynamically whenever the action is exact indicating the dynamical stability of geometric space. The model is intended as a setting for subsequent studies of emergent time mechanisms.
\end{abstract}

%
%
%
%
%

\section{Introduction}
One basic and widespread intuition in quantum gravity is that spacetime should be described by discrete variables, since this automatically regularises any gravitational theory: see e.g. section IV of \cite{Oriti_AppQG}. In loop quantum gravity \cite{Rovelli_QG} this discreteness is fundamental, arising spontaneously in the spectra of area and volume operators due to the compactness of the holonomy group $SU(2)$ \cite{RovelliSmolin_DiscAreaVolume}. Classical smooth spacetimes then occur as the continuum limit of certain spin network states known as \textit{weaves} \cite{AshtekarRovelliSmolin}. In causal set theory \cite{Sorkin_CST, Henson_CSAQG}, discreteness remains fundamental, but seems to be more an expression of the ontological (or at least operational) incoherence of the infinite \cite{EllisMeissnerNicolai_PhysInf}. For the dynamical triangulations programme \cite{AJL, AJL_QG, AGJL}, discreteness is an effect of a UV cutoff as in lattice field theory; the discreteness scale $\ell$ is then sent to $0$, and if this can be done in a cutoff independent way the theory has a well-defined continuum limit. Finally, the situation in string scenarios is rather subtle. At one level, string theory admits a fundamental length scale \cite{GrossMende_STBeyPS, Witten_ReflectionsFateSpacetime}, which should regularise the theory and give spacetime a granular structure. On the other hand, quantum effects might `smear out' discrete spacetimes into continua \cite{Martinec_CCC}, where the fundamental length obstructs any probes of spacetime discreteness. 

Let $\Omega$ denote a family of \textit{discrete} structures; typical examples of $\Omega$ include locally finite posets (causal set theory), piecewise linear structures (dynamical triangulations) and matrices (matrix models); in the approach adopted here $\Omega$ will consist of a family of random regular graphs. The basic problem of emergent geometry is to specify some (dynamical) selection principle which picks out an $\omega\in \Omega$ that approximates a solution of Einstein gravity in the continuum limit. One popular option (c.f. e.g. \cite{HanHung_LQG_RT, ChircoEtAl, Glaser_FinSizeScale, AJL_QG, Martinec_CCC, Bianconi}) is to exploit the formalism of statistical mechanics \cite{LandauLifshitz_SP, Goldenfeld_SM}: consider the (formal) partition function
\begin{eqnarray}\label{equation:PartitionFunction}
\mathcal{Z}=\sum_{\omega\in \Omega} e^{-\beta \mathcal{A}_{DEH}(\omega)},
\end{eqnarray}
where $\beta$ is some parameter of the theory and $\mathcal{A}_{DEH}$ is a discrete Einstein-Hilbert action with dimensions $[\beta^{-1}]$. Given an entropy $S=-\sum_{\omega\in \Omega} p(\omega) \log p(\omega)$, Boltzmann probability $p(\omega)=\exp(-\beta \mathcal{A}_{DEH}(\omega))/\mathcal{Z}$ and expectation $\braket{f}:=\sum_{\omega\in \Omega}f(\omega)p(\omega)$ for any function $f:\Omega\rightarrow \mathbb{R}$, we have an associated free energy
\begin{eqnarray}\label{equation:FreeEnergy}
F=\braket{\mathcal{A}_{DEH}}-\frac{S}{\beta}.
\end{eqnarray}
In the limit $\beta\rightarrow \infty$, equilibrium states minimise the expected action and we identify $\beta\rightarrow \infty$ as the classical limit. Later we shall identify
\begin{eqnarray}
\beta=(\hbar g)^{-1}
\end{eqnarray}
where $g$ is the gravitational coupling strength in units of the Newtonian gravitational constant. We thus see that the thermodynamic classical limit $\beta\rightarrow \infty$ agrees with the standard classical limit $\hbar\rightarrow 0$. We denote the equilibrium phase as $\beta\rightarrow \infty$ by $\Omega_{\infty}$, and call this the \textit{geometric} or \textit{classical} phase. The problem of emergent geometry thus reduces to demonstrating $(\omega,\mathcal{A}_{DEH}(\omega))\rightarrow (\mathcal{M},\mathcal{A}_{EH}(\mathcal{M}))$ for any $\omega\in \Omega_{\infty}$ where $\mathcal{M}$ is a smooth manifold and $\mathcal{A}_{EH}$ is the Einstein-Hilbert action. Any $\omega\in \Omega$ which does approximate a smooth manifold will be called a \textit{discrete manifold}, and we denote the set of all discrete manifolds by $\Omega_{ST}$. Clearly we require $\Omega_{\infty}\subseteq \Omega_{ST}$.

It is worth mentioning in passing that gauge-gravity duality \cite{HorowitzPolchinski, Nishimura_STOMM} and its generalisations such as the Ryu-Takayanagi conjecture \cite{RyuTakayanagi, Qi_EHM}---which states that the entanglement entropy of a conformal field theory in region $A$ of the boundary of some AdS space is proportional to the area of $\gamma_A$, a so called \textit{minimal surface} which is cobordant and action minimising---provide an alternative paradigm for emergent spacetime. Here gravitational theories are found to emerge from (conformal) field theories defined on the boundary of spacetime via some holographic duality. When the boundary theory is discrete (a lattice theory or matrix model, for instance) this formalism provides an attractive framework for thinking about emergent spacetime. Rather interestingly, the Ryu-Takayanagi duality can be understood in terms of conventional statistical mechanics if one interprets averages over random tensor networks as Ising model partition functions; then treating domain walls as Ryu-Takayanagi minimal surfaces ensures that the entanglement entropy satisfies the appropriate area law \cite{HaydenEtAl}. This tensor network description both brings the AdS/CFT (Ryu-Takayanagi) formalism for emergent spacetime much closer to the statistical formalism adopted here, but also grants scope for this formalism to be realised in nonperturbative background independent scenarios \cite{HanHung_LQG_RT, ChircoEtAl}.

Returning to the statistical formalism for emergent spacetime, we note that when the discrete Einstein-Hilbert action \eref{equation:PartitionFunction} is Euclidean, the partition function specifies a discretisation of a \textit{Euclidean} quantum gravity theory \cite{Hawking_EQG}. There are two interpretations of such a Euclidean formulation. The first is to consider the partition function as a Wick-rotated Lorentzian path integral \cite{GlimmJaffe}, and the statistical analysis acts as an effective sum-over-histories. This presents sizeable technical advantages since it permits the application of powerful computational methods from statistical mechanics \cite{NewmanBarkema, ADJ}. There are also several advantages from the perspective of mathematical rigour, such as the definition of path integrals via Wiener measures. In this viewpoint, time is fundamental and the Euclidean formulation is just a technical help: end results have to be Wick-rotated back to Lorentzian signature. An alternative point of view is that Euclidean quantum gravity is fundamental and time is an emergent concept that has to be `derived' from the model, possibly only at large scales. This is the point of view adopted here. Statistical analysis, however, continues to play an independent key role as long as spacetime discreteness is not fundamental. In particular, the divergence of the correlation length (treated as an observation scale) associated with any continuous phase transition allows one to take $\ell\rightarrow 0$ where $\ell$ is an effective discreteness scale (cutoff). From a renormalisation group (RG) perspective, the continuous phase transition defines {Einstein gravity as an IR fixed point of the theory; one alternatively views this as non-perturbative evidence for the existence of a UV completion of gravity in line with the asymptotic safety scenario \cite{Weinberg_AsymSaf}. Either way, the presence of a continuous phase transition indicates that the regularisation of gravity implied by spacetime discreteness is cutoff independent and the implied fixed point defines a consistent continuum theory at all scales.

Let us consider the `energy'-entropy balance \cite{Goldenfeld_SM} of the system. First let $\Omega_N\subseteq \Omega$ denote the set of all discrete spacetime structures with $N$ spacetime points. If $\mathcal{A}_{DEH}$ is local, then it may be expressed as a sum over spacetime points of some bounded function and the action is bounded above by $\alpha N$ where $\alpha$ is some finite positive constant. We thus see that each $\omega\in \Omega_N$ contributes at least $\exp(-\alpha N)$ to the partition function, implying $p(\omega)>0$ and $S\sim \log|\Omega_N|$. Thus if $|\Omega_N|$ has faster than exponential growth with $N$, the entropy term of the free energy \eref{equation:FreeEnergy} dominates for finite $\beta$. This leads to two potential problems: first, if $|\Omega_{ST}|\ll|\Omega|$ then a typical graph in the equilibrium phase will not admit a continuum limit. Secondly, the system has a homogeneous phase structure and no phase transition will be observed. In \cite{BenincasaDowker_ScalCurvCausSet}, this argument is used to justify the non-locality of the causal set action which, in essence, is defined as the number of short causal chains in the causal set. Non-locality in the action then expresses the fact that underlying causal relations are not necessarily local. Dynamical triangulations, on the other hand, use a local action defined via the Regge calculus. The old Euclidean approach ran afoul of the first problem, viz. the equilibrium phase at high $\beta$ consisted of baby universes which were not discrete manifolds. \textit{Causal }dynamical triangulations improve on this situation by restricting the configuration space $\Omega$. 

An alternative solution to the problem of excess entropy, which in particular does not make use of any causal structure, is to exploit the additional richness of statistical mechanics in networks \cite{AlbertBarabasi_StatMechCompNet, ParkNewman_StatMech}, apropos ordinary statistical mechanics. In conventional statistical systems such as the Ising model, the basic paradigm is the emergence of long range order from purely local interactions: for any bulk spin the set of possible local interactions is fixed irrespective of the size of the system and as long as boundary terms become negligible in the thermodynamic limit, the critical behaviour of the system will be independent of system size. Concretely this means that the phase of the system depends only on the relative \textit{value }of some parameter of the system with respect to a critical value. In random graph models, however, edges take the role of local interactions, and the set of possible local interactions depends on the global structure of the system. In particular as the system size $N\rightarrow \infty$, the number of possible local interactions of a vertex also diverges and we say random graph models are \textit{infinite dimensional systems}. The phase of the system thus depends on the comparison of some parameter with some critical \textit{function} which depends on system size. In effect, rather than compensating for excess entropy by introducing non-local terms in the action as in causal set theory, we utilise the intrinsically non-local nature of statistical mechanics in networks: local dynamics in a non-local space, so to speak. This has the obvious advantage that the action remains local in the process. 

Practically speaking, we renormalise the system parameter $\beta\mapsto \tilde{\beta}(\beta,N)$ in order to cancel any excess $N$ dependence in the entropy; we interpret this as the scale dependence of the quantum gravitational coupling $\hbar g$. This essentially factors out any $N$ dependence in the critical function and we expect to observe a critical value for the renormalised coupling. The precise dependence of $\tilde{\beta}$ on $N$ emerges as a consistency constraint on the formal continuum limit of the action and leads to area-law scaling for the entropy. Note that in principle, one could also derive the $N$-dependence of $\beta$ via an analysis of finite size scaling; an example is \cite{Glaser_FinSizeScale}, which finds several results that are rather akin to our own. Unfortunately, the authors have found such an analysis impractical in the present model.

Following previous work by one of the authors (see \cite{Trugenberger_QGasNetSO, Trugenberger_CSPN, Trugenberger_RHLW, DiamantiniTrugenber_TopNetEnt} and especially \cite{Trugenberger_CombQG}), this paper presents an approach to Euclidean quantum gravity in which geometry dynamically emerges from random regular graphs via a continuous phase transition driven by a local discretisation of the Einstein-Hilbert action. No Lorentzian or causal structure is needed in the process. In contrast to simplicial gravity approaches which use the Regge calculus, the discrete Einstein-Hilbert action is specified via the \textit{Ollivier curvature} \cite{Ollivier_RCMCMS, Ollivier_RCMS}, a \textit{local} discretisation of the manifold Ricci curvature defined for graphs augmented with a canonical metric measure structure. Locality comes from the fact that the Ollivier curvature of an edge is totally specified by the network structure of a known \textit{core neighbourhood} of the edge. Much of the utility of the Ollivier curvature in discrete settings arises from the fact that it admits rather explicit formulations in terms of purely combinatorial variables for certain classes of graphs \cite{JostLiu_RicciCurv, ChoPaeng_RicCurvCol, BhattacharyaMukherjee, LoiselRomon_RicciCurvPolySurf}, though its definition in terms of optimal transport theory \cite{Villani_OptimalTransport} makes it a burdensome drain on computational power when exact combinatorial expressions are unknown \cite{SamalEtAl}. We present a new exact result \eref{equation:NewOllivCurv} for the Ollivier curvature of an edge in terms of the number of short cycles (to be explained fully below) supported on that edge. This expression should be of independent interest to network theorists, firstly as the only exact expression for the Ollivier curvature for graphs with more than one type of short cycle, and secondly as an extension of previous results \cite{DallChristensen, Krioukov} relating clustering to geometric structure. 

The class of graphs for which the new result is valid is defined by a network analogue of a statistical mechanical hard core condition; effectively the hard core condition states that particles (short cycles) in the graph may touch (share an edge) but not overlap (share more than an edge). This condition is central to the neat combinatorial expression for the Ollivier curvature in terms of numbers of short cycles. At a physical level, we shall find that the hard core condition is both key for the dynamical suppression of defects appearing in the discrete manifold structure and is sufficient to prevent classical (action minimising) solutions from crumpling into baby universes, a traditional problem with the Euclidean dynamical triangulations framework. It thus plays a stabilising role similar to hard bosonic cores in the theory of Bose condensates or Pauli blocking in ordinary matter. The correct analogue depends on whether short cycles are to be regarded as `bosonic' or `fermionic'; there are no clear indications on this matter as yet. Taking a more `gravitational' perspective on the hard core condition, it can be seen as a restriction to the (combinatorially) simplest non-trivial configuration space in which the overall sign of the scalar curvature is not specified \textit{a priori}. The hard core condition thus defines a kind of minimal `solvable' model with potentially non-trivial realisations. Dynamically, however, it is found that flat spacetimes are favoured in the classical limit and the hard core condition can be seen as a restriction to irrelevant perturbations about a flat spacetime background. Weakening this condition is thus closely related to the problem of incorporating a cosmological constant or matter into the model; it is also desirable for the sake of making the model background independent. We reflect on how realistic such a procedure is in the conclusion.

Concretely, close analysis of the discrete Einstein-Hilbert action defined via the Ollivier curvature in a configuration space consisting of random regular graphs subject to the hard core condition suggests that classical solutions consist of cubic complexes up to a small and possibly negligible number of defects. Specialising to the bipartite case removes defects and ensures that $\Omega_\infty$ consists of graphs locally isomorphic to subsets of $\mathbb{Z}^n$. Such graphs have a natural interpretation as discrete Ricci flat manifolds. A mean field approximation for this case had previously been studied in \cite{Trugenberger_CombQG}; we extend the results of \cite{Trugenberger_CombQG}, giving the first evidence for the self-assembly of discrete manifolds from random regular graphs as well as direct evidence for the existence of a \textit{continuous} phase transition at finite $\beta$ in the form of a diverging correlation length plot. The stabilising role of the hard core condition becomes apparent when we study the system in mean field approximation: vacua now consist of baby universes with too many short cycles per edge, configurations which are dynamically excluded by the exact action. All numerical simulations are based on Monte Carlo Metropolis algorithms \cite{NewmanBarkema}. 

We would like to finish this introduction with a brief and rather heuristic discussion of the picture of (Euclidean) quantum gravity that this model entails. Unsurprisingly, given that the system is a statistical (field) theory, our basic paradigm is the Ising model. The main difference with the standard Ising model is that the fundamental degrees of freedom in our system are edges (connections) instead of vertices (points) and so we are forced to consider the statistical mechanics of networks, with the incumbent difficulties of infinite dimensionality discussed above. Although edges are the fundamental degrees of freedom on which the model is based, the `physical' degrees of freedom are network cycles; this expresses a kind of gravitational gauge principle. Physical degrees of freedom (cycles or loops) are local, in the sense that they are short (short cycles are either triangles, squares or pentagons). The core neighbourhood of an edge then consists of the neighbours of the points in the edge and the short loops based on that edge; this corresponds essentially to the set of nearest neighbours in the Ising model. From this perspective the hard core condition ensures that our physical degrees of freedom are well-separated, just as spins in the Ising model or electromagnetic field lines in lattice gauge theories. At high temperatures, distinct Ising spins have independent dynamics due to the large entropy, so that a typical configuration is random. Correspondingly, at strong coupling, the dynamics of each edge in our network becomes random, and spacetime becomes a random graph. Decreasing the coupling, we observe a second-order network phase transition, and the (square) cycle degrees of freedom `condense', which amounts to a self-organization of the graph into a regular lattice constituting the discrete approximation of a flat geometric manifold. At this critical point, the diverging correlation length allows the discretisation to be removed. Exactly as in the Ising model, and in line with the asymptotic safety scenario, this ultraviolet (UV) critical point thus provides a non-perturbative definition of (Euclidean) quantum gravity. The important new point is that, in our model, this UV completion of quantum gravity is defined by a \textit{network critical point} which has the immediate consequence that the entropy scales according to the famed area law. 
\section{The Ollivier Curvature in Graphs}
As mentioned above, the Ollivier curvature is a discretisation of the Riemannian Ricci curvature,and is defined using ideas from optimal transport theory. It is one of a series of rough curvatures appearing in metric measure geometry  \cite{Najman_DiscreteCurvature}, but it is particularly suited to \textit{discrete} settings and has recently attracted attention in applied network theory as a measure of network properties such as clustering and robustness \cite{Ni_RicIntTop, Sandhu_Cancer, Sandhu_Market}. Moreover, the Ollivier curvature of an edge is discrete, bounded and local in the sense that the curvature of an edge only depends on short cycles supported by that edge, where a cycle is short if its length is at most 5. This makes it an attractive model of a quantised gravitational field in some lattice regularisation; indeed, a somewhat similar proposal for a `quantum Ricci curvature' closely related to the Ollivier curvature in its basic intuition is made in \cite{KlitgaardLollI, KlitgaardLollII} from a dynamical triangulations perspective.

One of the main drawbacks of the Ollivier curvature is that it is difficult to compute in arbitrary graphs, and recent empirical studies \cite{SamalEtAl, Tannenbaum_BiologyComp}, have instead looked at correlations between the Ollivier curvature and alternative, more readily computable, discretisations of the Ricci curvature. There has been a line of research, however, seeking to give exact expressions for the Ollivier curvature in certain classes of graph. Ollivier \cite{Ollivier_RCMCMS} gave an exact result for the Ollivier curvature of an edge in $\mathbb{Z}^n$ ($\mathbb{Z}^n$ is Ollivier-Ricci flat) but the first non-trivial exact result in graphs seems to be due to Jost and Liu \cite{JostLiu_RicciCurv} who gave an exact expression for the Ollivier curvature of trees and complete graphs; these essentially represent the extreme points at which certain curvature inequalities are tight. Jost and Liu also emphasised the connection between Ollivier curvature and clustering; in view of other studies \cite{DallChristensen, Krioukov} on the connection between clustering and geometry we expect there to be links between Ollivier curvature and network geometry. Cho and Paeng \cite{ChoPaeng_RicCurvCol} slightly generalise the result on trees to graphs of girth at least 6, where the two expressions are in fact equivalent due to the locality property discussed above (graphs of girth at least $6$ are locally tree-like). These results were significantly extended by Bhattacharya and Mukherjee in \cite{BhattacharyaMukherjee} to graphs of girth at least $5$ and to bipartite graphs. Loisel and Romon \cite{LoiselRomon_RicciCurvPolySurf} demonstrate that the Ollivier curvature depends only on the degree of the vertices for edges in a surface triangulation satisfying certain genericity assumptions and consequently calculate the Ollivier curvature for many polyhedral surfaces. 

We continue this line of research by presenting a new exact expression for the Ollivier curvature in graphs satisfying a hard core condition. The hard core condition plays an important heuristic role in simplifying the basic mathematical procedure used to deduce exact expressions for the Ollivier curvature, in particular by reducing the core neighbourhood of each edge to the simplest form that includes all three types of short cycle. It also ensures that the edge curvature can be expressed purely in terms of the numbers of short cycles supported on the edge. Despite the importance of the hard core condition for the derivation of the result, the authors would like to stress that they believe the main result \eref{equation:NewOllivCurv} is essentially generic as an exact expression of the Ollivier curvature, at least in its qualitative features. Also as mentioned above, the hard core condition is sufficient for the stability of discrete manifolds as classical solutions of the system as will be shown in section \ref{section:IR}.
\subsection{Basic Properties of the Ollivier Curvature} \label{subsection:OlivCurvI}
We begin by briefly recalling some elementary features of the Ollivier curvature in a general metric measure space. More complete discussions of some of the relevant notions are available in chapters 1--6 of \cite{Villani_OptimalTransport} and \cite{Ollivier_RCMCMS}. Let $(X,\rho)$ be a complete metric space and suppose that we have a family of probability measures $\set{\mu_x}_{x\in X}$ indexed by $X$ defined on the Borel $\sigma$-algebra $\Sigma_X$ of $X$. This data defines a \textit{metric measure space}. Any two points $x,\:y\in X$ define a pair of probability spaces $(X,\Sigma_X,\mu_x)$ and $(X,\Sigma_X,\mu_y)$. A \textit{coupling} of the probability spaces $(X,\Sigma_X,\mu_x)$ and $(X,\Sigma_X,\mu_y)$ is a probability space $(X^2,\Sigma_{X^2},\pi_{xy})$ such that $\mu_x$ and $\mu_y$ are the pushforward measures of $\pi_{xy}$ under the projections $(x,y)\mapsto x$ and $(x,y)\mapsto y$ respectively. Equivalently, for any measurable function $f:X\rightarrow \mathbb{R}$ we have
\numparts
\begin{eqnarray}
\int_{X^2}\rmd\pi_{xy} (u,v)f(u)=\int_X\rmd\mu_x(u) f(u)\label{equation:MarginalConstraint1}\\
\int_{X^2}\rmd\pi_{xy} (u,v)f(v)=\int_X\rmd\mu_y(v) f(v).\label{equation:MarginalConstraint2}
\end{eqnarray}
\endnumparts
$\pi_{xy}$ is called a \textit{transport plan} between $\mu_x$ and $\mu_y$, which in turn are called the \textit{marginals} of $\pi_{xy}$. We denote the family of all transport plans between $\mu_x$ and $\mu_y$ by $\Pi(x,y)$. 

To each transport plan $\pi_{xy}$ between $\mu_x$ and $\mu_y$ we associate a \textit{transport cost:}
\begin{eqnarray}\label{equation:TransportCost}
W^\pi(\mu_x,\mu_y)=\int_{X^2}\rmd\pi_{xy}(u,v)\rho(u,v).
\end{eqnarray}
The basic intuition is that $\mu_x$ determines a distribution of a finite volume of dirt about the point $x$ while $\mu_y$ determines a hole of the same volume about the point $y$. Since the volumes are the same we normalise the total volume to $1$. $W^\pi(\mu_x,\mu_y)$ then determines the total cost of transporting the dirt distributed about $x$ to the hole distributed about $y$ where the cost of moving the dirt at a point $u$ near $x$ into the hole at a point $v$ near $y$ is the distance $\rho(u,v)$ between the two points. The \textit{Wasserstein distance} $W$ is a metric function on the space $\set{\mu_x}_{x\in X}$ of probability measures associated to $X$ defined by the expression
\begin{eqnarray}
W(\mu_x,\mu_y)=\inf_{\pi_{xy}\in \Pi(x,y)}W^\pi(\mu_x,\mu_y).
\end{eqnarray}
It is thus the \textit{optimal} cost of transport. The Ollivier curvature is then:
\begin{eqnarray}\label{equation:OllivCurv}
\kappa(x,y)=1-\frac{W(\mu_x,\mu_y)}{\rho(x,y)}.
\end{eqnarray}
From this it is clear that the Ollivier curvature is entirely determined given $W$ and $\rho$.

One fundamental feature of the Wasserstein distance is that it admits a dual formulation in terms of \textit{$1$-Lipschitz functions}. Recall that a $1$-Lipschitz function, also called a \textit{short map}, is a mapping $f:X\rightarrow Y$ where $(X,\rho_X)$ and $(Y,\rho_Y)$ are metric spaces such that $\rho_Y(f(x),f(y))\leq \rho_X(x,y)$ for all $x,\:y\in X$. The class of all short maps between $X$ and $Y$ is denoted $\mathcal{L}_1(X,Y)$. To every short map we associate the \textit{transport profit}:
\begin{eqnarray}\label{equation:TransportProfit}
W^f(\mu_x,\mu_y)=\int_X \rmd\mu_x(u) f(u)-\int _X\rmd\mu_y (u)f(u).
\end{eqnarray}
The Kantorovich duality theorem then states that
\begin{eqnarray}
W(\mu_x,\mu_y)=\sup_{f\in \mathcal{L}_1(X,\mathbb{R})}W^f(\mu_x,\mu_y).
\end{eqnarray}
An immediate consequence of this theorem is that
\begin{eqnarray}
W^f(\mu_x,\mu_y)\leq W(\mu_x,\mu_y)\leq W^\pi(\mu_x,\mu_y)
\end{eqnarray}
for any short map $f:X\rightarrow \mathbb{R}$ and any transport plan $\pi_{xy}\in \Pi(x,y)$. In particular if we can specify a transport plan $\pi_{xy}$ and a short map $f:X\rightarrow \mathbb{R}$ such that $W^\pi(\mu_x,\mu_y)=W^f(\mu_x,\mu_y)$ then $W^\pi(\mu_x,\mu_y)=W^f(\mu_x,\mu_y)=W(\mu_x,\mu_y)$. This will be our basic strategy for finding exact expressions of the Ollivier curvature.

Before moving to a discrete setting it is worth considering the Ollivier curvature in an ordinary Riemannian manifold. In particular we now assume that $X$ is an $D$-dimensional Riemannian manifold and that $\rho(x,y)$ is induced by the metric tensor as the length of any geodesic between $x$ and $y\in X$. Then for each $\ell>0$ we can define a probability measure at each point $x\in X$ by the assignment
\begin{eqnarray}
\rmd\mu_x^\ell(y)=\left\{\begin{array}{rl}
\frac{\rmd\rm{vol}(y)}{\rm{vol}[B_\ell(x)]}, & y\in B_\ell(x)\\
0, & y\notin B_\ell (x)
\end{array}\right.
\end{eqnarray}
where $B_\ell(x)$ denotes the open ball of radius $\ell$ centred at $x$. This makes $X$ into a metric measure space. If we now consider a pair of points $x,\:y\in X$ such that $y$ lies on the geodesic defined by some unit tangent vector $v$ at $x$ and such that $\rho(x,y)$ is sufficiently small, it can be shown that
\begin{eqnarray}\label{equation:OllivCurvMan}
\kappa(x,y)=\frac{\ell^2}{2(D+2)}\rm{Ric}(v,v)
\end{eqnarray}
up to higher powers in small terms, where $\rm{Ric}$ is the Ricci curvature tensor on $X$. The intuition is that for a positively curved manifold, the average distance between two small balls is smaller than the distance between their centres.
\subsection{Consequences of the Discrete Setting}\label{subsection:OlivCurvII}
In this section we demonstrate how the discrete setting allows for the deduction of exact expressions for the Ollivier curvature in terms of combinatorial variables. The basic ideas were first presented in \cite{JostLiu_RicciCurv} though the method was refined in \cite{BhattacharyaMukherjee}. Using the methods of \cite{BhattacharyaMukherjee}, we then present a new exact result for the Ollivier curvature in a wide class of graphs. Notation and terminology for this section is summarised in \ref{appendix:Notation}, and is specified in light of \cite{Diestel_GraphTheory}. 

We consider a (simple) graph $\omega=(V(\omega),E(\omega))$. Every (connected) graph $\omega$ is made into a metric space by defining the metric function $\rho_\omega$ such that $\rho_\omega(u,v)$ is the length of any minimal path between $u$ and $v$. Clearly $\rm{Im}(\rho_\omega)\subseteq \mathbb{N}$. Moreover if the graph is locally finite, i.e. if the degree $d_\omega(u)<\infty$ for each $u\in V(\omega)$, one can define a probability measure $\mu_u$ for each $u\in V(\omega)$ such that---using random walk terminology---nearest neighbour transitions have uniform probability $1/d_\omega(u)$ and all other transitions are suppressed. Thus locally finite connected graphs define metric measure spaces. Henceforth we assume that every graph is locally finite and connected unless specified otherwise.

There are several simplifications that immediately occur due to the discrete setting. Firstly any connected graph $\omega$ has a discrete metric topology and so its Borel $\sigma$-algebra is simply given by the power set of $V(\omega)$. Moreover each measure $\mu_u$ on $V(\omega)$ has finite support since the graph is locally finite and thus by countable additivity it is uniquely determined by the vector $\bmu_u=d_\omega(u)^{-1}1_{d_\omega(u)}$ where $1_n$ denotes an $n$-dimensional (real) column vector with all entries equal to $1$. More significantly, any transport plan $\pi_{uv}$ is uniquely determined by some $d_\omega(u)\times d_\omega(v)$ dimensional matrix ${\bpi_{uv}}$ satisfying the following constraints:
\numparts
\begin{eqnarray}
{\bpi_{uv}}1_{d_\omega(v)}={\bmu_u}\label{equation:DiscreteMarginalConstraint1}\\
{\bpi_{uv}}^T1_{d_\omega(u)}={\bmu_v}.\label{equation:DiscreteMarginalConstraint2}
\end{eqnarray}
\endnumparts
Conversely any such matrix specifies a unique transport plan $\pi_{uv}\in \Pi(u,v)$ on $V(\omega)^2$. Equations \eref{equation:DiscreteMarginalConstraint1} and \eref{equation:DiscreteMarginalConstraint2} should be interpreted as the discrete expression of the marginal constraints \eref{equation:MarginalConstraint1} and \eref{equation:MarginalConstraint2} respectively. In fact it should be clear that matrices of the appropriate dimensions satisfying the discrete marginal constraint conditions \eref{equation:DiscreteMarginalConstraint1} and \eref{equation:DiscreteMarginalConstraint2} specify essentially identical transport plans on all spaces $X^2$ where $N_\omega(u)\cup N_\omega(v)\subseteq X\subseteq V(\omega)$. Note that $N_\omega(x)$ denotes the neighbours in $\omega$ of any $x\in V(\omega)$. We will not distinguish between these transport plans since they differ only by sets of null measure in the domain.

Given a transport plan $\pi_{uv}$, we may write down a discrete expression for the transport cost \eref{equation:TransportCost}:
\begin{eqnarray}
W_\omega^\pi(\mu_u,\mu_v)={\bpi_{uv}}\cdot \bi{D_{\bomega}^{uv}}
\end{eqnarray}
where the \textit{distance matrix} $\bi{D^{uv}_{\bomega}}$ is a $d_\omega(u)\times d_\omega(v)$ dimensional matrix such that $\bi{D^{uv}_{\bomega}}_{(x,y)}=\rho_\omega(x,y)$ for any $(x,y)\in N_\omega(u)\times N_\omega(v)$ and $\cdot$ denotes the element-wise inner product, i.e. the inner product of ${\bpi_{uv}}$ and $ \bi{D^{uv}_{\bomega}}$ regarded as $d_\omega(u)\cdot d_\omega(v)$-dimensional column vectors. Thus calculating the transport cost of a transport plan becomes an elementary exercise in linear algebra. Moving on to the Kantorovich dual formulation, it is sufficient to note that in the discrete setting the transport profit of any bounded map $f:V(\omega)\rightarrow \mathbb{R}$ is
\begin{eqnarray}\label{equation:DiscreteTransportProfit}
W_\omega^f(\mu_u,\mu_v)=\frac{1}{d_\omega(u)}\sum_{w\in N_\omega(u)}f(w)-\frac{1}{d_\omega(v)}\sum_{w\in N_\omega(v)}f(w).
\end{eqnarray}
For short maps, \eref{equation:DiscreteTransportProfit} above agrees with the general expression for the transport profit \eref{equation:TransportProfit}, where we have utilised the fact that the supports of $\mu_u$ and $\mu_v$ are $N_\omega(u)$ and $N_\omega(v)$ respectively.

The next two simplifications that arise in the discrete setting have important physical implications. The first is the discreteness of the Wasserstein metric. In particular $\rm{Im}(W_\omega)\subseteq \mathbb{Q}$, a consequence of the fact that:
\begin{eqnarray}
\sup_{f\in \mathcal{L}_1(X,\mathbb{R})} W_\omega^f(\mu_u,\mu_v)=\sup_{f\in \mathcal{L}_1(X,\mathbb{Z})} W_\omega^f(\mu_u,\mu_v).
\end{eqnarray}
This essentially states that it is sufficient to optimise over integer-valued $1$-Lipschitz maps. This may be proven using some well known results from linear programming theory \cite{BhattacharyaMukherjee}. The second simplification of physical significance is the locality property discussed above. To formulate this, we introduce the notion of a \textit{core neighbourhood}: a connected graph $C_{uv}\subseteq \omega$ such that $N_\omega(u)\cup N_\omega(v)\subseteq V(C_{uv})\subseteq V(\omega)$ is a \textit{core neighbourhood} of $u$ and $v\in V(\omega)$ iff $\bi{D_{C_{uv}}^{uv}}_{(x,y)}\leq  \bi{D_{\bomega}^{uv}}_{(x,y)}$ for all $(x,y)\in N_{\omega}(u)\times N_\omega(v)$. As $C_{uv}$ is a subgraph of $V(\omega)$ by definition, $\rho_\omega(x,y)\leq \rho_{C_{uv}}(x,y)$ for any pair of vertices $x,\:y\in V(C_{uv})$; this follows because every path in $C_{uv}$ is a path in $\omega$. Thus $C_{uv}$ is a core neighbourhood of $u$ and $v\in V(\omega)$ iff $\bi{D_{C_{uv}}^{uv}}=  \bi{D_{\bomega}^{uv}}$. Recalling that a matrix ${\bpi_{uv}}$ satisfying the constraints \eref{equation:DiscreteMarginalConstraint1} and \eref{equation:DiscreteMarginalConstraint2} uniquely specifies identical transport plans on all sets $X$ such that $N_\omega(u)\cup N_\omega(v)\subseteq X\subseteq V(\omega)$, we immediately see that
\begin{eqnarray}
W_\omega(\mu_u,\mu_v)=W_{C_{uv}}(\mu_u,\mu_v)
\end{eqnarray}
for any pair of points $u,\:v\in V(\omega)$, where subscripts denote the graph on which the Wasserstein metric is defined.

This is already a locality property, and it only remains to make it more explicit by specifying a core neighbourhood. For an edge $uv\in E(\omega)$ this is very simple. The key point to note is that between any two points $x\in N_\omega(u)$ and $y\in N_\omega(v)$ there is a $3$-path $xuvy$ and $\rho_\omega(x,y)\leq 3$. This immediately implies that the induced subgraph $c_{uv}\subseteq \omega$ defined by the vertex set $V(c_{uv})=N_\omega(u)\cup N_\omega(v)\cup \pentagon(uv)$ is a core neighbourhood of $u,\:v\in V(\omega)$ where $\pentagon(uv)$ is the set of all points that are a distance $2$ from both $u$ and $v$. This is because any path between $x$ and $y$ not in $c_{uv}$ is of length at least $3$ and there is a path at least as short in $c_{uv}$. If the distance between two distinct points $(x,y)\in N_\omega(u)\times N_\omega(v)$ is less than $3$ then we may adjoin the shortest path between these two points to obtain a square or a pentagon supported on the edge $uv$. If $x=y$ then $x$ clearly defines a triangle supported by $uv$. If the distance is $3$ then we have a shortest path $xuvy$ and so the elements of the core neighbourhood are neighbours of $u$ and $v$ or lie on a short cycle supported by the edge $uv$ where a cycle is short iff its length is less than $6$. $c_{uv}$ will be called the \textit{standard core neighbourhood} of $uv\in E(\omega)$.

Henceforth we shall only consider the Wasserstein distance and Ollivier curvature at an edge $uv\in E(\omega)$: 
\begin{eqnarray}
\kappa_\omega(uv)=1-W_\omega(\mu_u,\mu_v).
\end{eqnarray}
Since we wish to define a local action in terms of the Ollivier curvature this is sufficient for our purposes.

\begin{figure}
	\centering
	\includegraphics[width=0.6\textwidth]{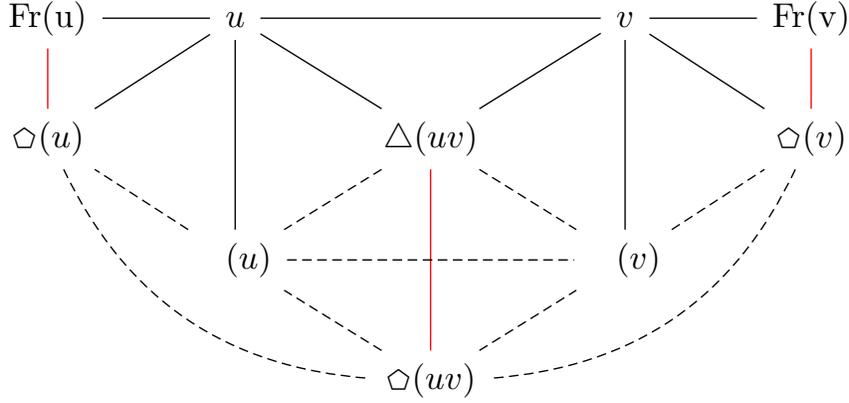}
	\caption{The core neighbourhood of an arbitrary edge in a random graph. The symbols appearing in the figure are defined in \ref{appendix:Notation} and form a partition of the core neighbourhood. Solid black lines denote necessary connections between the partition elements, while dashed lines denote possible connections. Red lines denote possible connections that can be deleted without changing the form of any block form distance matrix with blocks given by the partition elements.}\label{figure1}
\end{figure}
The standard core neighbourhood of an edge $uv\in E(\omega)$ is represented diagrammatically in \fref{figure1}. The symbols appearing in the figure are defined in \ref{appendix:Notation}, but we have tried to choose a relatively self-evident notation: $\triangle$, $\square$ and $\pentagon$ all denote sets of core neighbourhood elements that lie on triangles, squares and pentagons respectively. The bracketed terms denote the closest of the vertices $u$ or $v$ to the set in question. The only remaining point of ambiguity are the symbols $\rm{Fr}(u)$ and $\rm{Fr}(v)$ which denote the neighbours of $u$ and $v$ respectively that do not lie on any short cycle. We call such vertices \textit{free} in the subsequent. Solid black lines between two subsets $X,\:Y\subseteq V(\omega)$ denote that every vertex in $X$ is adjacent to a vertex in $Y$ and vice versa; dashed lines denote that a vertex in $X$ is possibly adjacent to a vertex in $Y$. Red lines denote potential vertices which can be deleted without changing the distance matrix.  Core neighbourhoods become a powerful heuristic tool whenever the representation only contains solid black lines as this immediately specifies a block form distance matrix $\bi{D^{uv}_{c_{uv}}}$. We present the transport plan matrices in a similar block form, weighing the row and column sums by the size of the block to recover the marginal constraints \ref{equation:DiscreteMarginalConstraint1} and \eref{equation:DiscreteMarginalConstraint2}. These ensure that every entry in a transport plan matrix $\bpi_{uv}$ lies in the interval $[0,1]$ and we thus obtain optimality via transport that prefers to move earth between nearby blocks.

We consider graphs satisfying a network analogue of the \textit{hard core conditions} which appear in ordinary statistical mechanics. More formally, an edge $uv\in E(\omega)$ satisfies the hard core condition iff any two short cycles supported on $uv$ share no other edges. We also say that the edge has \textit{independent short cycles}. A \textit{graph} satisfies the hard core condition if all edges have independent short cycles. Equivalently, the hard core condition says that the abstract graphs presented in \fref{figure2} are excluded subgraphs.
\begin{figure}
	\centering
	\begin{subfigure}{0.3\textwidth}
		\centering
		\includegraphics[width=0.5\textwidth]{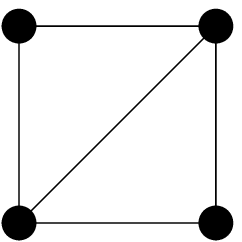}
		\caption{$(\triangle,\square)$}\label{figure2a}
	\end{subfigure}
	\begin{subfigure}{0.3\textwidth}
		\centering
		\includegraphics[width=0.5\textwidth]{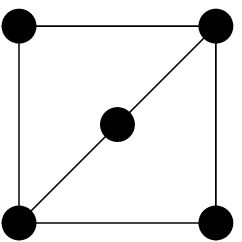}
		\caption{$(\square,\square)$}\label{figure2b}
	\end{subfigure}
	\begin{subfigure}{0.3\textwidth}
		\centering
		\includegraphics[width=0.5\textwidth]{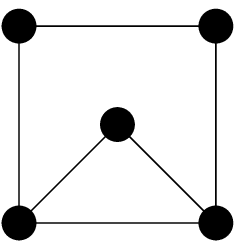}
		\caption{$(\triangle,\pentagon)$, $(\square,\pentagon)$}\label{figure2c}
	\end{subfigure}
	\begin{subfigure}{0.3\textwidth}
		\centering
		\includegraphics[width=0.5\textwidth]{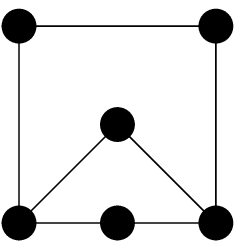}
		\caption{($\square,\pentagon$), $(\pentagon,\pentagon)$}\label{figure2d}
	\end{subfigure}
	\begin{subfigure}{0.3\textwidth}
		\centering
		\includegraphics[width=0.5\textwidth]{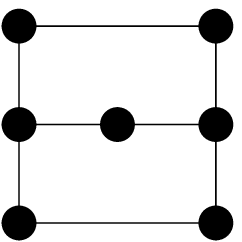}
		\caption{$(\pentagon,\pentagon)$}\label{figure2e}
	\end{subfigure}
	\caption{Excluded subgraph characterisation of the hard core condition. Listed pairs for each subgraph give the types of short cycle sharing more than one edge.}\label{figure2}
\end{figure}
The independent short cycle condition is an obstruction to sharing more than a single edge, so we intuitively treat short cycles as particles which can touch (share an edge) but not overlap (share multiple edges). This will have important physical ramifications described in section \ref{section:IR}, but for present purposes we can justify this condition by its effect on the standard core neighbourhood as displayed in figure \ref{figure3}.
\begin{figure}
	\centering
	\includegraphics[width=0.6\textwidth]{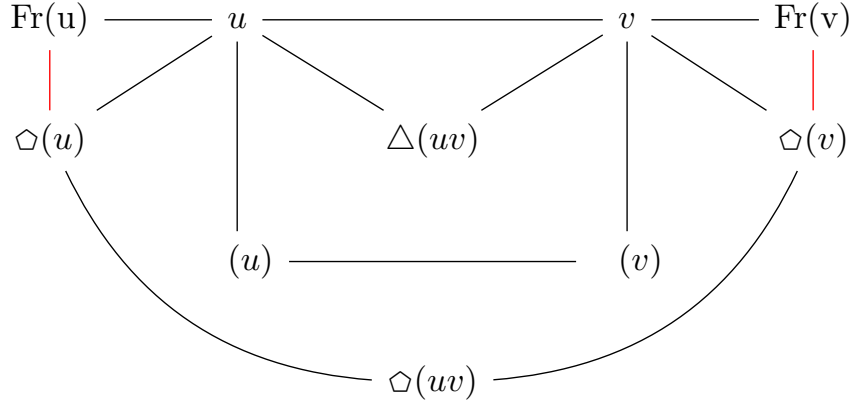}
	\caption{The core neighbourhood of an edge with independent short cycles; this representation of the core neighbourhood partition immediately specifies a block form distance matrix.}\label{figure3}
\end{figure}
The distance matrix is then:
\begin{eqnarray}\label{equation:DistanceMatrix}
\begin{array}{c|ccccc}
\bi{D_{c_{uv}}^{uv}} & u & \triangle(uv) & \square(v) & \pentagon(v) & \rm{Fr}(v)\\
\hline
v & 1 & 1 & 1 & 1 & 1\\
\triangle(uv) & 1 & 0 & 2 & 2 & 2\\
\square(u) & 1 & 2 & 1 & 3 & 3\\
\pentagon(u) & 1 & 2 & 3 & 2 & 3\\
\rm{Fr}(u) & 1 & 2 & 3 & 3 & 3
\end{array}.
\end{eqnarray}
It is worth stressing that a more refined partition may allow for the specification of a similar block form matrix given a weaker assumption than independent short cycles.

Using this matrix as a heuristic guide it becomes possible to specify \textit{optimal} transport plans. These can be used to derive our \textit{main equation:}
\begin{eqnarray}\label{equation:NewOllivCurv}
\fl \kappa(uv)=\frac{\triangle_{uv}} {d_u\land d_v}-\left[1-\frac{1+\triangle_{uv} +\square_{uv}}{d_u\lor d_v}-\frac{1}{ d_u\land d_v}\right]_+\nonumber\\
- \left(\frac{\triangle_{uv}}{d_u\land d_v}-\frac{\triangle_{uv}}{d_u\lor d_v}\right)\lor \left(1-\frac{1+\triangle_{uv} +\square_{uv} +\pentagon_{uv}}{d_u\lor d_v}-\frac{1}{d_u\land d_v}\right).
\end{eqnarray}
$d_u:=d_{\omega}(u)$ and $d_v:=d_\omega (v)$. $\triangle_{uv}$, $\square_{uv}$ and $\pentagon_{uv}$ denote the number of triangles, squares and pentagons supported on the edge $uv$ respectively. $n_u$ and $n_v$ denote the number of free neighbours of $u$ and $v$ respectively. $\alpha\lor\beta:=\max(\alpha,\beta)$, $\alpha \land\beta:=\min(\alpha,\beta)$ and $[\alpha]_+:=0\lor\alpha $ for any $\alpha,\:\beta\in \mathbb{R}$. The full proof of \eref{equation:NewOllivCurv} is given in \ref{appendix:Proof}. Remarkably, the hard core condition allows us to calculate the Ollivier curvature by simply counting the number of short `loops' on an edge.

We finish this section by discussing our heuristic belief that the hard core condition is essentially generic, despite its technical nature. First note that if a core neighbourhood does not have independent short cycles, then either it contains elements that lie on distinct types of short cycles that share a pair or more of edges or it contains elements that lie on two or more of the \textit{same} type of short cycle that share more than two edges. In the first case, let us consider a configuration isomorphic to \fref{figure2a} for the sake of concreteness. An optimal transport plan will see the top right vertex as an element of a $3$-cycle; the cost of this is that we are forced to treat the top left vertex as an effectively free vertex and we must perturb the number of free vertices in the final expression of edges lying on distinct types of short cycle. In practice the situation is a little more involved, but the basic idea holds. Configurations where an edge lies on two or more of the same type of cycle that share at least two edges have already been dealt with in \cite{BhattacharyaMukherjee} by taking sums over connected subgraphs. These two modifications can be made without fundamentally changing the combinatorial expression for the Ollivier curvature and in this sense we say that the hard core condition is generic.
\section{Structure of the Geometric Phase}\label{section:IR}
In this section we analyse various structural features of the equilibrium or geometric phase in the formal classical limit $\beta\rightarrow 0$. Equivalently, one can see this section as an analysis of the IR limit of the theory. There are three key features of this section: firstly, a specification of the dynamics of the model via the presentation of an exact action (c.f. section \ref{subsubsection:GeneralDynamics}). Secondly, in section \ref{subsection:DiscreteManifolds}, we give a (relatively) precise characterisation of the family $\Omega_{Disc}^D$ of discrete $D$-manifolds. Finally, in sections \ref{subsection:OddCycleDefect} and \ref{subsection:SqDefects}, we demonstrate that the \textit{vacua} or \textit{classical solutions} (action minimising configurations) of the theory are (at least approximately) discrete manifolds.
\subsection{Discrete Manifolds}\label{subsection:DiscreteManifolds}
In this section we specify the dynamics of the theory and characterise the discrete manifolds. The dynamics is given by a natural Euclidean generalisation of the Einstein-Hilbert action when viewing the Ollivier curvature as a rough Ricci curvature. Discrete manifolds are then abstractly characterised by a series of formal limits which are required in order to obtain the `correct' continuum limit of the theory. Regularity emerges as a kinematic constraint in this context, which, however, in conjunction with the hard-core condition, essentially proves to be sufficient to stabilise discrete manifolds as the classical solutions of the system. Demonstrating this is the purpose of sections \ref{subsection:OddCycleDefect} and \ref{subsection:SqDefects}.
\subsubsection{General Dynamics.} \label{subsubsection:GeneralDynamics}
We define the discrete Einstein-Hilbert action via the assignment
\begin{eqnarray}\label{equation:Action1}
\mathcal{A}_{DEH}(\omega)=-\sum_{u\in V(\omega)}d_u\sum_{v\in N_\omega (u)}\kappa_\omega(uv).
\end{eqnarray}
The idea is that we have an action
\numparts
\begin{eqnarray}
\mathcal{A}_{DEH}(\omega)=-\sum_{u\in V(\omega)}d_u\kappa_\omega(u)\\
\kappa_\omega(u):=\sum_{v\in N_\omega (u)}\kappa_\omega(u,v).
\end{eqnarray}
\endnumparts
That is to say, we have a local action defined as a weighted sum over the total edge curvature $\kappa_\omega(u)$ at a vertex $u\in V(\omega)$. For all cases considered in this paper, the weighting $d_u$ will reduce to an overall constant and is simply introduced to match the conventions of \cite{Trugenberger_CombQG}, which are used in some of the later numerical simulations. We do not expect this to cause any qualitative change to the critical behaviour of the system, though we do expect it to have an effect on the value of the critical coupling.

We are interested in the continuum limit $\beta\rightarrow \infty$ of the dimensionless action $\beta \mathcal{A}_{DEH}$ where $\beta=(\hbar g)^{-1}$. First define a discrete manifold as any graph $\omega\in \Omega$ which admits the following limits:
\numparts
\begin{eqnarray}
\omega\rightarrow\mathcal{M} \label{equation:FormalReplacements1}\\
d_u\sum_{v \in N_\omega(u)}\kappa(u,v)\rightarrow \frac{D\ell^2}{2(D+2)}R\label{equation:FormalReplacements2}\\
\sum_{u\in V(\omega)}\ell^D\rightarrow \int_{\mathcal{M}}\rm{d vol}_D\label{equation:FormalReplacements3} 
\end{eqnarray}
\endnumparts
where $\mathcal{M}$ is a Riemannian manifold, $\rm{d vol}_D$ is the volume element of $\mathcal{M}$, $R$ is the scalar curvature and $\ell\rightarrow 0$ is an effective discreteness scale. Heuristically this definition is rather natural: the first limit is trivial. The second limit essentially follows from \eref{equation:OllivCurvMan} assuming that the Ollivier curvature is preserved in the limit. Note that we have implicitly assumed regularity for reasons discussed in section \ref{subsubsection:RegConstraint} below. Given that the metric is diagonal and Riemannian, the volume element is simply an infinitesimal hypercube and the final limit is also trivial. It should be clear that $\Omega_{ST}$ contains subsets of $\mathbb{Z}^D$, though it is still not entirely clear if non-trivial discrete manifolds exist.

Treating $|V(\omega)|$ as a dimensionless universe volume and writing $\ell=\ell_0 |V(\omega)|^{-1/D}$ where $\ell_0$ is a fixed length scale (a cosmological scale) identifies the continuum limit $\ell\rightarrow 0$ and the thermodynamic limit $|V(\omega)|\rightarrow \infty$. Finally, assuming that $\omega$ is a discrete manifold we make the replacements \eref{equation:FormalReplacements1}, \eref{equation:FormalReplacements2} and \eref{equation:FormalReplacements3} to obtain:
\begin{eqnarray}
\frac{1}{\hbar g}\mathcal{A}_{DEH}\rightarrow -\frac{1}{2(D+2)\ell_0^{D-2}}\frac{|V(\omega)|^{1-\frac{2}{D}}}{\hbar g}\int_{\mathcal{M}} \rm{dvol}_D R
\end{eqnarray}
which is formally a Euclidean version of the vacuum Einstein-Hilbert action. However we see that the RHS diverges in the thermodynamic limit for $D>2$ unless we renormalise $g$ and allow it to grow as $|V(\omega)|^{1-2/D}$. We thus identify $\tilde{\beta}=(\hbar g)|V(\omega)|^{2/D-1}$ and investigate the phase transition in terms of $\tilde{\beta}$. Recalling the free energy $F=\braket{\mathcal{A}_{DEH}}-S/\tilde{\beta}$, and the growth $\braket{\mathcal{A}_{DEH}}\sim |V(\omega)|$ we see that energy-entropy balance immediately implies that
\begin{eqnarray}
|S|\sim |V(\omega)|^{\frac{2}{D}}
\end{eqnarray}
Since $|V(\omega)|$ is the dimensionless volume of the universe, $|S|$ thus scales as an area ensuring that we have recovered an area-entropy law.

We consider a configuration space $\Omega$ of graphs $\omega$ with $|V(\omega)|=N$ and $|E(\omega)|=E$ constant. Recall the free energy $F=\braket{A_{DEH}}-S/\tilde{\beta}$ where we have replaced the inverse gravitational coupling $\beta$ by a renormalised inverse coupling; equilibrium states minimise the free action and define the geometric phase $\Omega_{ST}$ in the classical limit $\tilde{\beta}\rightarrow \infty$. Thus one can understand the geometric phase, at least to a first approximation, by considering graphs that minimise the expected action. Since the continuum limit of action \eref{equation:Action1} is the (Euclidean) Einstein-Hilbert action by construction, matter is absent from the model and equilibrium states discretely approximate solutions to the vacuum Einstein field equations. All such continuum solutions are Ricci flat, so we expect the equilibrium phase to consist of Ollivier-Ricci flat graphs.
\subsubsection{Regularity, Discrete Manifolds and Defects.}\label{subsubsection:RegConstraint}
The limit \eref{equation:FormalReplacements2} used in characterising discrete manifolds essentially imposes regularity as a constraint on the discrete manifolds. In particular, to interpret $R$ as the trace of $\rm{Ric}$, we need to treat each edge intersecting with $u$ as a basis vector of length $\ell$ in the tangent plane of $u$, where the metric is diagonal in this basis. Since the tangent space at a point on a manifold must have the dimension of that manifold this essentially forces us to consider $D$-regular graphs, although if we can pair `antiparallel' edges with identical curvatures then $2D$-regular graphs remain an option. Taking subsets of $\mathbb{Z}^D$ as the basic paradigm of discrete manifolds, we expect $2D$-regular graphs to be of primary interest. Regularity is a highly non-perturbative constraint, and in lieu of any indications that regularity appears spontaneously in the system, we must impose regularity as a constraint from the outset. Thus we consider the configuration space $\Omega=\Omega_{d,N}$, the space of $d$-regular graphs satisfying the hard core condition on $N$ vertices.

In this configuration space, we can find a useful reformulation of the action. First note that the curvature is given
\begin{eqnarray}
\kappa(uv)=\frac{\triangle_{uv}} {d}-\left[1-\frac{2+\triangle_{uv} +\square_{uv}}{d}\right]_+ - \left [1-\frac{2+\triangle_{uv} +\square_{uv} +\pentagon_{uv}}{d}\right]_+.
\end{eqnarray}
Substituting into the expression \eref{equation:Action1} we obtain
\begin{eqnarray}
\fl \mathcal{A}(\omega)=\left(2Nd(d-2)-18\triangle_{\omega}-16\square_{\omega}-10\pentagon_{\omega}\right)-\sum_{uv\in P_\omega}\left((d-2)-(\triangle_{uv}+\square_{uv})\right)\nonumber\\
-\sum_{uv\in Q_\omega}\left((d-2)-(\triangle_{uv}+\square_{uv}+\pentagon_{uv})\right)
\end{eqnarray}
where we define
\numparts
\begin{eqnarray}
P_\omega:=\set{uv\in E(\omega):(\triangle_{uv}+\square_{uv})>(d-2)}\\
Q_\omega:=\set{uv\in E(\omega):(\triangle_{uv}+\square_{uv}+\pentagon_{uv})>(d-2)}.
\end{eqnarray}
\endnumparts
The first term is obtained by performing the sum of $\kappa(uv)$ ignoring the subscript $+$, while the latter two terms are required to compensate for this fact. Note that
\begin{eqnarray}
\triangle_{\omega}=\frac{1}{6}\sum_{u\in V(\omega)}\sum_{v \in N_\omega(u)}\triangle_{uv}\qquad 
\square _{\omega}=\frac{1}{8}\sum_{u\in V(\omega)}\sum_{v\in N_\omega(u)}\square_{uv}\qquad  \pentagon_{\omega}=\frac{1}{10}\sum_{u\in V(\omega)}\sum_{v\in N_\omega(u)}\pentagon_{uv}.
\end{eqnarray}
The factors of $6=2\cdot 3$, $8=2\cdot 4$ and $10=2\cdot 5$ in the denominators come from the fact that the sum $\sum_{u\in V(\omega)}\sum_{v\in N_\omega(u)}$ sums over all edges $uv\in E(\omega)$ twice, while each triangle/square/pentagon has 3/4/5 edges. 

We define the combinatorial variables:
\begin{eqnarray}\label{equation:LocalCombFields}
\varphi^\omega_\triangle(uv):=\frac{\triangle_{uv}}{d-2}\qquad  \varphi^\omega_\square(uv):=\frac{\square_{uv}}{d-2}\qquad  \varphi^\omega_{\pentagon}(uv):=\frac{\pentagon_{uv}}{d-2}
\end{eqnarray}
These variables simply count the number of short cycles supported on an edge under some normalisation, the significance of which will only become apparent later. Such variables define, however, a combinatorial analogue of local one-form fields and we can thus introduce their mean field approximations:
\begin{eqnarray}
\phi_{\triangle}(\omega):=\braket{\varphi_{\triangle}^\omega(uv)}_{E(\omega)}\qquad \phi_{\square}(\omega):=\braket{\varphi_{\square}^\omega(uv)}_{E(\omega)}\qquad \phi_{\pentagon}(\omega):=\braket{\varphi_{\pentagon}^\omega(uv)}_{E(\omega)}
\end{eqnarray}
where $\braket{\cdot}_{E(\omega)}$ denotes that the average is taken with respect to the edges of $\omega$. Explicitly, we find:
\begin{eqnarray}
\phi_\triangle(\omega)=\frac{\triangle_{\omega}}{\frac{1}{6}Nd(d-2)}\qquad \phi_\square(\omega)=\frac{\square_{\omega}}{\frac{1}{8}Nd(d-2)}\qquad \phi_{\pentagon}(\omega)=\frac{\pentagon_{\omega}}{\frac{1}{10}Nd(d-2)}.
\end{eqnarray}
The denominators give the number of 3/4/5-cycles in a graph with $(d-2)$ 3/4/5-cycles per edge. Given that $Q_\omega\subseteq P_\omega$, we may now rewrite the action:
\numparts
\begin{eqnarray}
\mathcal{A}(\omega)=\mathcal{A}_{MF}(\omega)+\mathcal{A}_{P}(\omega)+\mathcal{A}_{Q}(\omega)\label{equation:RegularAction}\\
\mathcal{A}_{MF}(\omega)=Nd(d-2)\left(2-3\phi_\triangle(\omega) -2\phi_\square(\omega) -\phi_{\pentagon}(\omega)\right)\label{equation:RegActMF}\\
\mathcal{A}_{P}(\omega)=(d-2)\sum_{uv\in P_\omega-Q_\omega}(\varphi_\triangle^\omega(uv)+\varphi_\square^\omega(uv)-1)\label{equation:RegActP}\\
\mathcal{A}_Q(\omega)=(d-2)\sum_{uv\in Q_\omega}(2\varphi_\triangle^\omega(uv)+2\varphi_\square^\omega(uv)+\varphi_{\pentagon}^\omega(uv)-2)\label{equation:RegActQ}.
\end{eqnarray}
\endnumparts
Thus the action \eref{equation:RegularAction} is the sum of three terms, a \textit{mean field} term $\mathcal{A}_{MF}$ \eref{equation:RegActMF} which only consists of global variables, and two local correction terms $\mathcal{A}_P$ \eref{equation:RegActP} and $\mathcal{A}_Q$ \eref{equation:RegActQ}, defined purely in terms of the\textit{ local }combinatorial `fields' \eref{equation:LocalCombFields}.

One incidental feature of regularity is that short cycles become asymptotically sparse in the strong coupling limit \cite{Wormald_ModRRG}. The latter fact is significant for two reasons: firstly if short cycles become asymptotically sparse in the strong coupling limit, the `fields' $\varphi$ and $\phi$ all vanish in this limit and it becomes appropriate to treat these fields as order parameters for a phase transition. Sparseness of short cycles in the strong coupling limit also has a gravitational interpretation: given $\triangle_{uv}=\square_{uv}=\pentagon_{uv}=0$ we have
\begin{eqnarray}
\kappa_\omega(uv)=-2\left(1-\frac{1}{d_u}-\frac{1}{d_v}\right), \qquad \kappa_\omega(u)=-2\left(d_u-1-\sum_{v \in N_\omega(u)}\frac{1}{d_v}\right).
\end{eqnarray}
By removing short cycles, we will essentially remove any correlation of the degree random variables for nearby points, while long cycles do not effectively correlate degrees between any points since many different long cycles can issue from the same edges. Thus the values of $\kappa_\omega$ are uncorrelated at distinct points and we have the effective uncoupling of the gravitational field at distinct points in the graph. This is in line with the strong coupling limit in canonical quantisation scenarios \cite{Pilati_StrongCouplingQG}. Of course in a regular graph $\kappa_\omega(u)$ above is simply a constant.

Regularity proves to be a powerful constraint in other ways. We will find below that analysis of the action \eref{equation:RegularAction} for a $2D$-regular configuration space suggests that a typical vacuum (action minimising configuration) is a discrete manifold up to defects. Concretely, we argue that most edges in a typical vacuum will support $(2D-2)$ squares and no other cycles. Given such an edge there is a local isomorphism to $\mathbb{Z}^D$; configurations with every edge supporting $(2D-2)$ squares thus specify discrete $D$-manifolds, where the limits \eref{equation:FormalReplacements1} and \eref{equation:FormalReplacements3} are due to the local isomorphisms with $\mathbb{Z}^D$ and the limit \eref{equation:FormalReplacements2} follows trivially since such configurations are Ricci flat. Defects to this discrete manifold structure come in two types: \textit{odd cycle defects}, i.e. edges which support an odd short cycle, and \textit{square defects}, or edges which only support squares but do not support $(2D-2)$ squares. We consider each type of defect in turn.
\subsection{Odd Cycle Defects}\label{subsection:OddCycleDefect}
In section \ref{subsubsection:RegConstraint} above, it was argued that (Ricci flat) discrete manifolds consist of $2D$-regular graphs with $(2D-2)$ squares at each edge, while it was shown that the action \eref{equation:RegularAction} could be expressed as the sum of mean field and local constraint terms in a regular configuration space. Meanwhile, any edge that supported an odd cycle or $n\neq (2D-2)$ squares was defined as a \textit{defect}. In this section the aim is to demonstrate that odd cycle defects are dynamically suppressed and explain why we opt for $2D$-regular graphs rather than $d$-regular graphs for $d$ odd. The situation with square defects is a little more subtle, and is dealt with in isolation of odd cycle defects in section \ref{subsection:SqDefects}. 

The essence of the argument is quite simple: short cycles are sparse in random regular graphs , so initially the core neighbourhood of a typical edge will be dominated by the free vertices. The mean field term of the action promotes the formation of short cycles with triangles favoured most and pentagons least. There is thus a dynamical hierarchy which promotes the formation of triangles and squares at an edge. At the same time, there is a kinematical constraint arising due to the hard core condition which severely suppresses triangle defects. Thus we expect a typical core neighbourhood to be dominated by squares, and we expect the vacua to consist of (approximate) cubic complexes \cite{Ziegler_Polytopes}. Numerical evidence (\fref{figure4}) supports these conclusions. Note that the local correction terms of the action play essentially no role in this argument; this can be shown explicitly.

Consider the  mean field action \eref{equation:RegActMF} more closely:
\begin{eqnarray}
\mathcal{A}_{MF}(\omega)=Nd(d-2)\left(2-3\phi_\triangle(\omega) -2\phi_\square(\omega) -\phi_{\pentagon}(\omega)\right).\nonumber
\end{eqnarray}
It is monotonically decreasing in the mean field variables and naively we minimise the action by maximising the number of triangles, squares and pentagons in the graph respectively, with greatest relative gains obtained coming from triangles and least coming from pentagons as indicated by the coefficients in the action. This is the dynamical hierarchy of preferred short cycles. If one considers the excluded subgraphs specified by figures \ref{figure2a} and \ref{figure2c}, however, it is immediately apparent that in a graph with independent short cycles, a triangle can only share an edge with a pentagon. Regularity then implies that for each triangle on an edge we immediately freeze in a large number of free vertices and/or pentagons; in light of the relative gains to be made by turning pentagons into squares, this suggests that triangles and pentagons are suppressed in equilibrium configurations, and a typical discrete manifold in the geometric phase is approximately a cubic complex \cite{Ziegler_Polytopes}. 

We now consider the role of the correction term $\mathcal{A}_{P}(\omega)$ \eref{equation:RegActP}:
\begin{eqnarray}
\mathcal{A}_{P}(\omega)=(d-2)\sum_{uv\in P_\omega-Q_\omega}(\varphi_\triangle^\omega(uv)+\varphi_\square^\omega(uv)-1)\nonumber.
\end{eqnarray}
By the hard core condition $0<\varphi_{\triangle}^\omega(uv)$ implies $\varphi_{\square}^\omega(uv)=0$ while $\varphi_\triangle^\omega(uv)=\triangle_{uv}/(d-2)<1$ since $\triangle_{uv}\leq 1$ for any $uv\in E(\omega)$. Thus edges supporting triangles never contribute to this term and we can assume $\varphi_\triangle^\omega(uv)=0$. The maximum number of short cycles that can be supported on an edge is $(d-1)$ by regularity so the term above contributes non-trivially iff $\square_{uv}=(d-1)$. As the term contributes positively to the action, edges with $(d-1)$ squares are dynamically suppressed relative to these with fewer squares. Similar considerations for the term $\mathcal{A}_Q(\omega)$ \eref{equation:RegActQ} suggest that edges with $(d-1)$ squares and edges with (d-2) squares and a pentagon are dynamically suppressed; otherwise there is no local \textit{dynamical} constraint on the number of triangles or pentagons on an edge. In sum, if an edge has (d-2) squares then favoured configurations are those in which the remaining edges at $u$ and $v$ are free.

The above analysis thus suggests that equilibrium graphs consist of cubic complexes with $(d-2)$ squares per edge, with triangles and pentagons appearing as possibly negligible defects. The normalisation factor $(d-2)$ has two effects. Regarding edges as tangent vectors, cycles define surfaces intersecting with the edges in question and any two edges that lie on the same cycle are linearly independent. If there are (d-2) cycles on an edge $uv\in E(\omega)$ edge, then at each vertex $u,\:v\in uv$ there is a unique edge incident to the vertex which does not lie on any cycle supported on $uv$. This allows us to continue the vector $uv$ to other points and essentially defines a canonical connection on the tangent bundle of the (discrete) manifold. Also, it couples tangent vectors at a point suggesting that the dimension of the discrete manifold should be $D=d/2$. This of course requires $d\in 2\mathbb{Z}$. Accepting this dimensional assumption, we do indeed find the dynamical suppression of triangles and pentagons:  \fref{figure4} is a typical example of the quenched short cycle dynamics in $D=2$ and demonstrates the expected behaviour.
\begin{figure}[h]
	\centering
	\includegraphics[width=0.6\textwidth]{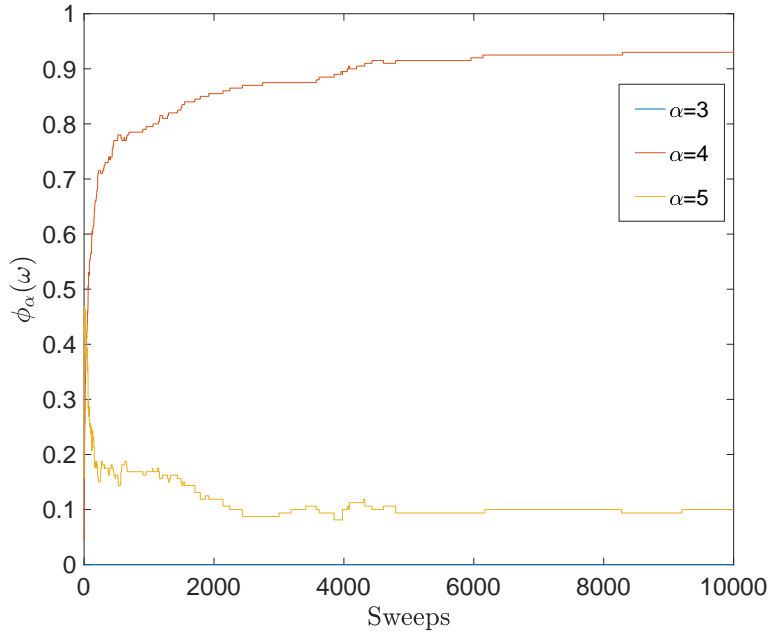}
	\caption{Quenched short cycle dynamics in $D=2$ for a graph with $N=200$ vertices. Note that $\phi_3(\omega):=\phi_{\triangle}(\omega)$, $\phi_4(\omega):=\phi_{\square}(\omega)$ and $\phi_5(\omega):=\phi_{\pentagon}(\omega)$. $\phi_{\square}(\omega)$ asymptotically approaches $1$, while $\phi_{\pentagon}(\omega)$ is dynamically suppressed. $\phi_{\triangle}(\omega)$ remains a vanishing constant throughout.}\label{figure4}
\end{figure}
\subsection{Square Defects}\label{subsection:SqDefects}
A bipartite graph is one in which there are no odd cycles; odd cycle defects are thus clearly absent from such graphs and the role of square defects is better studied in a bipartite configuration space. In fact, the arguments of the previous section suggest that this is a reasonable approximation whenever we are only interested in the bulk properties of our spacetime graphs. We also assume that $d=2D$ where $D$ is regarded as the manifold dimension.

For $2D$-regular bipartite graphs, the curvature may be expressed:
\begin{eqnarray}\label{equation:RegBipCurv}
\kappa_\omega(uv)=-2\left(1-\frac{1}{D}\right)\left[1-\varphi_{\square}^\omega(uv)\right]_+
\end{eqnarray}
and $\kappa(uv)\leq 0$ and $\mathcal{A}(\omega)\geq 0$ trivially, where equality holds iff $\omega$ is Ricci flat; we denote the set of $2D$-regular bipartite Ricci flat graphs by $\Omega_0$. From \eref{equation:RegBipCurv}, Ricci flat configurations are those with at least $2(D-1)$ squares on each edge. Noting that $P_\omega=Q_\omega$ for regular bipartite graphs, the action becomes
\numparts
\begin{eqnarray}
\mathcal{A}(\omega)=\mathcal{A}_{MF}(\omega)+\mathcal{A}_{\rm{cor}}(\omega)\label{equation:FullAction}\\
\mathcal{A}_{MF}(\omega)=8ND(D-1)(1-\phi_{\square}(\omega))\label{equation:MFA}\\ \mathcal{A}_{\rm{cor}}(\omega)=2\left(D-1\right)\sum_{uv\in P_\omega}(\varphi_{\square}^\omega(uv)-1).
\end{eqnarray}
\endnumparts 

As in the non-bipartite case, the action can be expressed as the sum of a mean field term and a local correction term, where the mean field term promotes the formation of short cycles (squares) and the local correction term suppresses them whenever there are more than $2D-2$ squares at an edge. Strictly speaking, the local correction term means that the global gain from the mean field term due to an increase in the total number of squares is precisely compensated for by the local correction term and the configurations are degenerate. Regardless, it is `energetically' favourable to increase the numbers of squares at an edge until that edge supports $2D-2$ squares, and then it becomes more profitable to increase the numbers of squares at another edge. This can continue until all edges support $2D-2$ squares. Hence a naive analysis of the `energetics' immediately implies that a square defect must support \textit{more} than $2D-2$ squares, and that the discrete manifold on $N$ vertices is a vacuum configuration. 

It is here that the crucial question of the \textit{stability} of the discrete manifold vacuum arises: every configuration with at least $2D-2$ squares at an edge is degenerate with the discrete manifold configuration so in order to obtain a consistent classical limit, these configurations must somehow be excluded. Essentially, the idea is that if we are in a geometric vacuum (discrete manifold) configuration, any local edge switch (Glauber transition) will either leave the graph unchanged or increase the number of squares at one edge by simultaneously decreasing the number of squares at another. The edge with a decreased number of squares will naturally increase the action; however, due to the local constraints, the edge with the increased number of squares will not lead to any reduction in the action. Thus distinct vacua lie in (topologically) distinct regions of configuration space, ensuring the stability of geometric space. Since a system initially in the random graph phase must enter a geometric vacuum state before it can become a non-geometric vacuum, we have the desired exclusion of non-geometric configurations.

The principal problem with this argument is that it does not hold in the thermodynamic limit. In particular the argument above relies on the dynamical significance of the local correction term; however since $\mathcal{A}_{MF}$ is manifestly extensive, the `correct' generalisation of the above argument depends strongly on the behaviour of $|P_\omega|$ as $N\rightarrow \infty$. We investigate this problem in section \ref{subsubsection:MeanField} below. For the purposes of this investigation it is helpful to identify the following:
\begin{enumerate}[label=(\Alph*)]
	\item $\omega\in \Omega_{A}:=\set{\omega\in \Omega: \phi_{\square}(\omega)<1}$. $\omega$ is not Ricci flat and $\Omega_A\cap \Omega_{0}=\emptyset$. For every $\omega\in \Omega_A$ there is an $\omega'\in \Omega_A$ such that $\phi_{\square}(\omega)=\phi_{\square}(\omega')$ and $\mathcal{A}_{\rm{cor}}(\omega')=0$. Thus $|P_\omega|<\infty$ as $N\rightarrow \infty$ for any action minimising graph in this regime.
	\item $\omega\in \Omega_{B}:=\set{\omega\in \Omega: \phi_{\square}(\omega)=1}$. Ricci flat configurations in $\Omega_B$ exist so $\Omega_B\cap\Omega_0\neq \emptyset$; such configurations must have $\varphi_{\square}^\omega(uv)=1$ for each $uv\in E(\omega)$. As $N\rightarrow \infty$, configurations have $|P_\omega|<\infty$.
	\item $\omega\in \Omega_C:=\set{\omega\in \Omega:\phi_\square(\omega)>1}$. Ricci flat configurations exist where each Ricci flat configuration in $\Omega_C$ has a non-zero fraction of edges satisfying $\varphi_{\square}^\omega(uv)>1$. As $N\rightarrow \infty$, $|P_\omega|$ also diverges.
\end{enumerate}
As we shall see, the argument for the dynamical exclusion of non-geometric vacua in section \ref{subsubsection:ExactDynamics} continues to rely on the significance of the local correction term $\mathcal{A}_{\rm{cor}}$ and is rather heuristic in its nature. Thus it becomes more important to consider numerical evidence for the dynamical stability of geometric space. One convenient way to consider the importance of the local correction term is by comparing the exact to the mean field dynamics. In doing so we find that
\begin{enumerate}
	\item Vacua of the mean field action \eref{equation:MFA} lie in $\Omega_C$. These vacua can be identified as the baby universes of Euclidean dynamical triangulations; see \fref{figure5} for an illustration.
	\item Possible vacua of the exact action \eref{equation:FullAction} lie in $\Omega_B\cup \Omega_C$. Only those vacua in $\Omega_B$ are dynamically realised.
\end{enumerate}
In particular discrete manifolds are not vacua in the mean field dynamics, and thus the `time' $\tau_{MF}$ taken for $\phi_{\square}(\omega)$ to exceed $1$ in the mean field dynamics specifies a stability timescale for $\Omega$. If, on the other hand, $\phi_{\square}(\omega)$ remains less than $1$ for a timescale $\tau\gg \tau_{MF}$, we note that the local correction term has significantly affected the dynamics, providing evidence for the (classical) stability of geometric space. We discuss these two cases in more detail below. Henceforth, unless specified otherwise $\phi(\omega):=\phi_{\square}(\omega)$.
\subsubsection{Mean Field Dynamics}\label{subsubsection:MeanField}
\begin{figure}[h]
	\centering
	\includegraphics[width=0.6\textwidth]{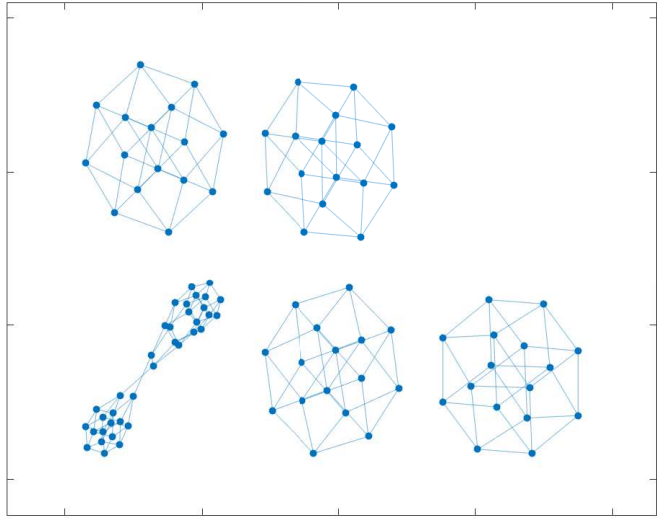}
	\caption{Baby universes that appear under the influence of the mean field action in $D=2$.}\label{figure5}
\end{figure}
Let $\Omega_{2D,N}$ denote the space of all $2D$-regular bipartite graphs on $N$ vertices satisfying the hard core condition. The mean field action \eref{equation:MFA} is monotonically decreasing in $\phi(\omega)$ which grows linearly with $\square_{\omega}$ for fixed $N$ and $D$. Thus the mean field action is minimised by maximising the number of squares in the graph. In a $2D$-regular graph with independent short cycles, it is clear that an edge can support at most $(2D-1)$ squares since given an edge $uv$, each of the $(2D-1)$ edges remaining at $u$ and $v$ are paired up uniquely by squares. From these considerations we immediately see that
\begin{eqnarray}
\square_{\omega}\leq ND(2D-1)
\end{eqnarray} 
for any $\omega\in \Omega_{2D,N}$. We say that $\omega$ is a \textit{baby universe} iff $\omega$ is connected and the above inequality is tight. For any baby universe $\omega_0$, $\phi(\omega_0)=\frac{2D-1}{D-1}$ and we have the following inequality for the intensive mean field action:
\begin{eqnarray}
\frac{1}{N}\mathcal{A}_{MF}(\omega_0)\geq -8D^2.
\end{eqnarray}
The RHS is the value of the intensive mean field action for any baby universe, and such configurations minimise the action. 

In the limit $N\rightarrow \infty$, configurations consisting of an infinite number of baby universes minimise the action \eref{equation:MFA}. The point is that given any baby universe $\omega_0$ on $M$ vertices, then for any $N\gg M$, since we have without loss of generality that $N=nM+r$ where $n$ and $r$ are natural numbers, there exists a configuration consisting of $n$ copies of $\omega_0$ as well as some configuration $\omega_r$ on the remaining $r$ vertices. This configuration has an intensive action of the form:
\begin{eqnarray}
\frac{1}{N}\mathcal{A}_{MF}(\omega)= -8D^2+\mathcal{O}\left(\frac{r}{N}\right),
\end{eqnarray}
where the second term vanishes when $\omega_r$ is square maximal. In the limit $N\rightarrow \infty$, the second term vanishes regardless and there is a vacuum configuration consisting of an infinite number of baby universes. This construction easily generalises to situations where more than one baby universe exists. We can also allow a finite number of components which are not baby universes

Thus configurations consisting of an infinite number of baby universes are mean field vacua. In fact we expect these vacua to be typical: the point is that for any vacuum configuration $\omega$ consisting of a finite number of connected components, one can construct an infinite number of configurations that include $\omega$ by appending any configuration $\omega_\infty$ which consists of an infinite number of baby universes. As long as there are `enough' configurations $\omega_{\infty}$ consisting of an infinite number of finite baby universes, the number of configurations with an infinite number of components will be infinitely richer than the number with only a finite number of components. Assuming that there is at most one baby universe on any $N$ vertices, and assuming that for any $N$ vertices there is a finite non-empty set of integers $n_1,...,n_k$ which are smaller than $N$ and support baby universes, the number of configurations $\omega_{\infty}$ is given by the limit as $N\rightarrow \infty$ of partitions of $N$ using the integers $n_1,..,n_k$ and $m$ other integers where $m\ll k$. In view of the very large number of partitions, there are probably enough configurations for the above considerations to be valid. This argument is of course rather heuristic and is best supported with additional evidence; in D=2 we do indeed typically observe configurations disconnecting into configurations consisting of several baby universes as shown in \fref{figure5}.

From the point of view of networks, it is found that graph connectivity is closely correlated to the existence of points of large negative curvature \cite{SamalEtAl}. For any graph $\omega$ with adjacency matrix $A$ \cite{HararyManvel},
\begin{eqnarray}
\phi_\triangle(\omega)\sim \Tr(A^3)\qquad \phi_\square(\omega)\sim \Tr(A^4) \qquad \phi_{\pentagon}(\omega)\sim \Tr(A^5)
\end{eqnarray}
so to a first approximation any action defined as the negative of the trace over a low order polynomial in an adjacency matrix will tend to maximise the Ollivier curvature and thus lead to the disconnection of graphs. It is possible that disconnected graphs are entropically suppressed for sufficiently high degree (dimension) due to the standard combinatorial result that random $d$-regular graphs are almost surely connected as $N\rightarrow \infty$ for $d\geq 3$. Note that $\Tr(A)$ vanishes while $\Tr(A^2)$ is simply a constant for regular graphs.
\begin{figure}[h]
	\centering
	\includegraphics[width=0.6\textwidth]{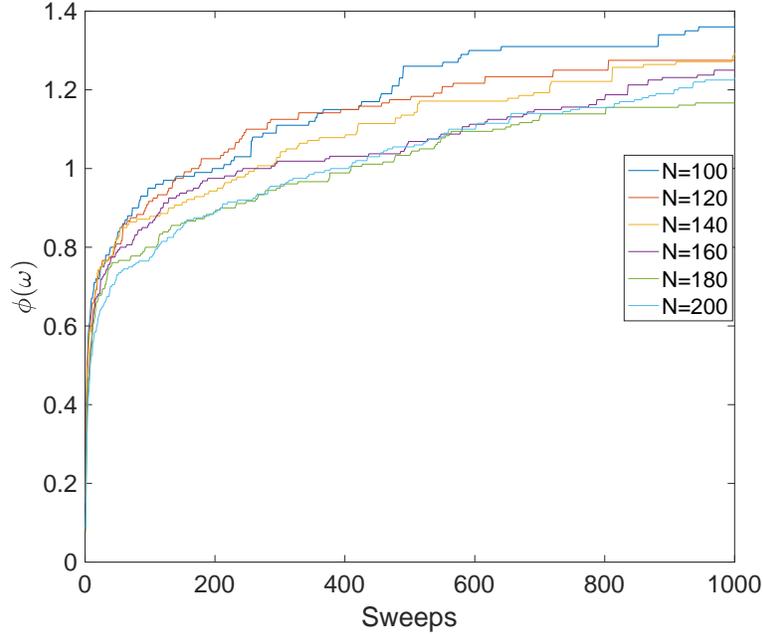}
	\caption{$\phi(\omega)$ for quenched mean field dynamics in $D=2$.}\label{figure6}
\end{figure}

We finish this section by briefly considering the mean field values of vacuum configurations under quenched dynamics: \fref{figure6} demonstrates that the vacua do indeed lie in the regime $\Omega_C$ and that baby universe configurations are in fact dynamically realisable. The timescale $\tau_{MF}\sim 1000$.
\subsubsection{Exact Dynamics}\label{subsubsection:ExactDynamics}
We now consider the dynamics of the exact action \eref{equation:FullAction}. By construction a configuration is a classical solution of this action iff it is Ricci flat. Thus in principle we have vacua in both regimes $\Omega_B$ and $\Omega_C$. Baby universes are also possible vacua. However, we find that the local correction term used in the full action dynamically suppresses vacuum configurations in the regime $\Omega_C$ since these have too many squares at each edge. We present both a theoretical argument and numerical evidence for this claim.

We argue that vacuum configurations in $\Omega_C$ are not dynamical solutions to the equations of motion: firstly, we note that for $\tilde{\beta}\rightarrow 0$, the system consists of random regular graphs. In such graphs, short cycles are sparse and the initial configuration $\omega_0\in \Omega_A$ almost surely. In particular $|P_{\omega_0}|<\infty$ even in the limit $N\rightarrow\infty$. The (Glauber) dynamics then defines a path from $\omega_0$ to some configuration $\omega_{\rm{vac}}$ where $\omega_{\rm{vac}}$ minimises the action \eref{equation:FullAction}. Since $|P_{\omega_0}|<\infty$, we expect $|P_{\omega_{\rm{vac}}}|<\infty$ and $\omega_{\rm{vac}}\in \Omega_B$. This justifies the claim as long as there is no subsequent transition between $\omega_{\rm{vac}}$ and some classical solution $\omega_{\rm{vac}}'\in \Omega_C$. For this point, however, it is sufficient to note that $|P_\omega|<\infty$ for any $\omega\in \Omega_B$ while $|P_\omega|\rightarrow \infty$ for any $\omega\in \Omega_C$ as $N\rightarrow \infty$; hence one requires an infinite number of Glauber transitions (edge switches) to go from any vacuum configuration in $\Omega_B$ to any vacuum configuration in $\Omega_C$. This is akin to the existence of an infinite energy barrier between the two configurations in the limit $N\rightarrow \infty$. 

Indeed, \fref{figure7} demonstrates that $\phi(\omega)=1$ acts as an (approximate) upper barrier to quenched configurations evolving under the exact action \eref{equation:FullAction} for $D=2$. In particular, approximate $\Omega_B$ vacua are stable for significantly longer run times under the exact action than it takes for a configuration to breach the barrier $\phi(\omega)=1$ under the mean field action \eref{equation:MFA}, as is readily apparent if one compares with \fref{figure6}. Note that the barrier $\phi(\omega)=1$ is only strict in the limit $N\rightarrow \infty$ and (observed) minor breaches are to be expected for smaller configurations.
\begin{figure}[h]
	\centering
	\includegraphics[width=0.6\textwidth]{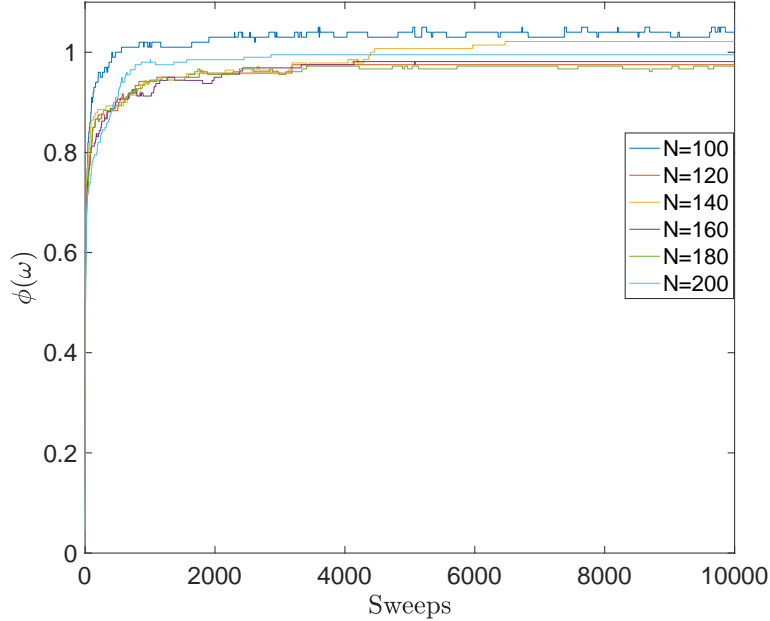}
	\caption{$\phi(\omega)$ for quenched dynamics under the exact action in $D=2$. The constraint $\phi(\omega)\approx 1$ arises dynamically.}\label{figure7}
\end{figure}
\section{From Random Regular Graphs to Discrete Manifolds}\label{section:Numerics}
In this section we present numerical evidence that there is a continuous phase transition at finite $\tilde{\beta}$ from a random phase of graphs to a geometric phase of discrete manifolds $\Omega_{Disc}^D$ in the sense of section \ref{subsection:DiscreteManifolds} (with an additional bipartiteness constraint). The phase transition seems to exist for both $D=2$ and $D=3$. Let $\Omega_{2D,N}$ denote the configuration space consists of $2D$-regular bipartite graphs with $N$ vertices and independent short cycles as above; then define $\tilde{\Omega}_{2D,N}:=\set{\omega\in \Omega_{2D,N}:P_\omega=\emptyset}$. We study the annealed dynamics of the mean field action \eref{equation:MFA} in the configuration space $\tilde{\Omega}_{2D,N}$. In this configuration space, both the exact action \eref{equation:FullAction} and the mean field action \eref{equation:MFA} agree, and the discrete manifolds $\Omega_{Disc}^D$ are automatically stable vacua. Since, as is argued in section \ref{subsubsection:ExactDynamics}, the constraint $P_\omega=\emptyset$ arises dynamically, it seems likely that the exact action on $\Omega_{2D,N}$ will drive a phase transition in the same universality class.

Note that \cite{Trugenberger_CombQG} considers the same model (though in $D=4$, rather than $D=2$). In \cite{Trugenberger_CombQG}, however, the constraint $P_\omega=\emptyset$ appeared implicitly via the numerics: results came from simulations of the annealed Glauber dynamics starting at \textit{high} $\tilde{\beta}$. The evident collapse of the value of $\phi(\omega):=\phi_{\square}(\omega)$ for graphs with distinct numbers of vertices provided good evidence for the existence of a two-phase structure and implied that the phase transition was independent of $N$ and thus continuous. However it failed to show the full self-assembly of discrete manifolds from random regular graphs, and no direct evidence for the continuous nature of the transition was presented. In contrast we anneal from low to high $\tilde{\beta}$ and demonstrate that random graphs at least approximately assemble into discrete manifolds belonging to $\Omega_{Disc}^D$; we also present a correlation length plot which displays clear divergent behaviour. The numerical results of this section can be regarded as extensions of findings in \cite{Trugenberger_CombQG}, and indeed this was the initial context in which they were gathered.

We first consider the two phase structure. 
\begin{figure}[h]
	\centering
	\begin{subfigure}{0.45\textwidth}
		\includegraphics[width=\textwidth]{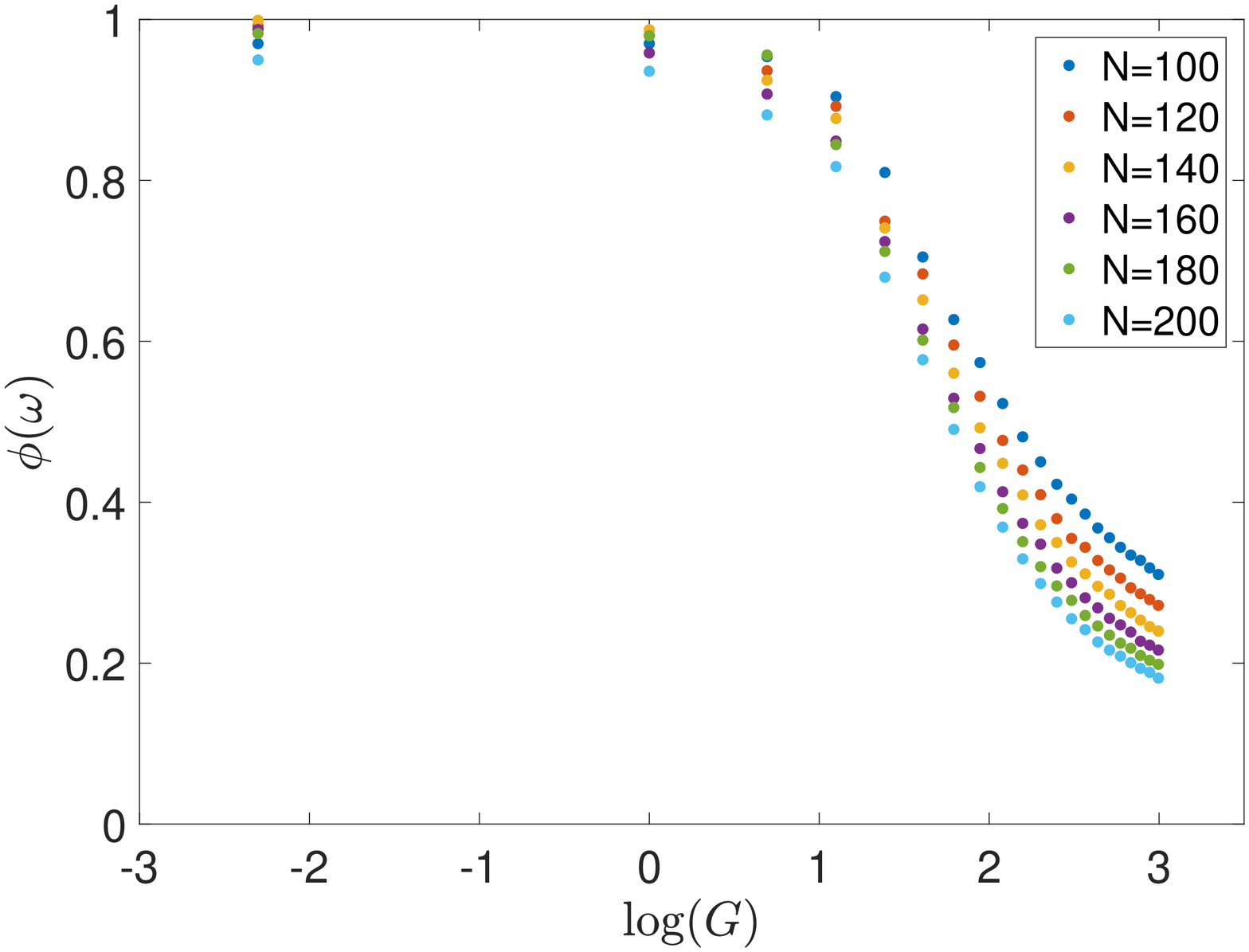}
		\caption{D=2.}\label{figure8a}
	\end{subfigure}
\begin{subfigure}{0.45\textwidth}
	\includegraphics[width=\textwidth]{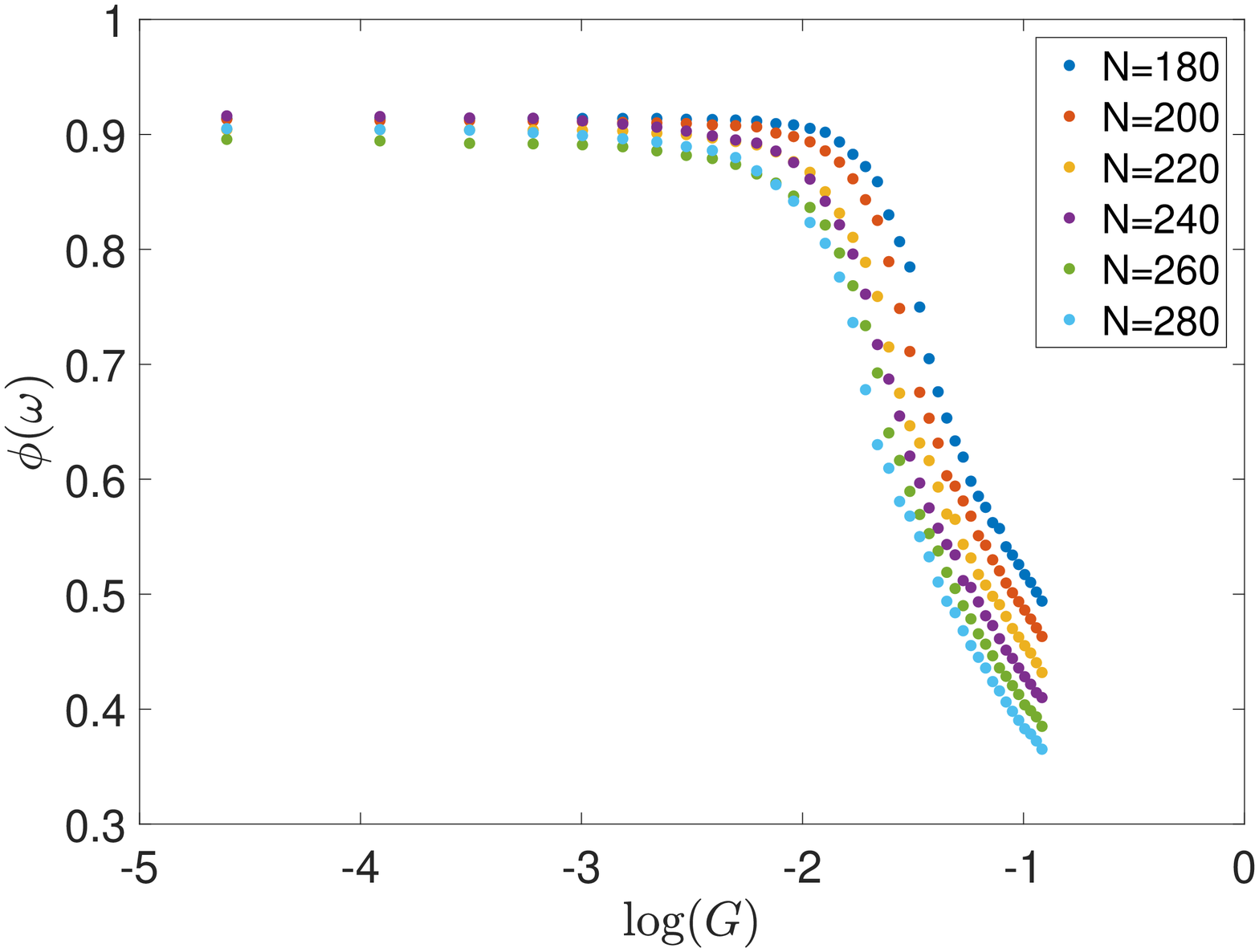}
	\caption{D=3.}\label{figure8b}
\end{subfigure}
	\caption{Variation of the mean field $\phi(\omega)$ with the ($\log$) inverse renormalised gravitational coupling $G=\tilde{\beta}^{-1}$ for $D=2$ and $D=3$. There is evidence in both dimensions for the existence of two distinct phases characterised by low ($\approx 0$) and high ($\approx 1$) values of $\phi(\omega)$ respectively. Note that in $D=2$, the geometric phase $\phi(\omega)=1$ is directly realised, whereas in $D=3$  finite size effects ensure that the geometric phase is only approximately realised.}\label{figure8}
\end{figure}
We see from \fref{figure8} that in $D=2$, the field $\phi(\omega)$ begins in $\Omega_A$ as discussed in section \ref{subsection:SqDefects} and approaches regime $\Omega_B$. Finite size effects prevent $\phi(\omega)$ from actually entering the discrete manifold phase since regular bipartite toroidal graphs are only supported on certain fixed values of $N$. Note that the sparseness of random regular graphs implies $\phi(\omega)\rightarrow 0$ as $N\rightarrow \infty$ making it an appropriate order parameter for the phase transition. It is worth noting that collapse of square density values is not evident until rather high values of $\tilde{\beta}$ are obtained, indicating that finite size effects are significant if there is indeed a second order phase transition.

Correlation lengths are a typical metric for the detection of continuous phase transitions. Specifically, we consider correlations between \textit{fluctuations} of the order parameter $\varphi(u):=\sum_{v\in N_\omega(u)}\varphi_{\square}(uv)/d$ about its mean value $\phi(\omega)$:
\begin{eqnarray}
C(u,v)=\frac{\braket{(\varphi(u)-\phi(\omega))(\varphi(v)-\phi(\omega))}_\omega}{\braket{(\varphi_{\square}(uv)-\phi(\omega))^2}_\omega}
\end{eqnarray}
where $\braket{\cdot }_\omega$ is expectation with respect to the graph, we can then formally define the correlation length
\begin{eqnarray}
\xi:=-\left\langle \frac{\rho(u,v)}{\log C(u,v)}\right\rangle_\omega.
\end{eqnarray}
The intuition is that as $\rho(u,v)\rightarrow \infty$, $C(u,v)\rightarrow \exp(-\rho(u,v)/\xi)$ whenever $G>G_c$, the critical coupling. Of course in finite graphs, especially random regular graphs, we are highly constrained in the maximum value of $\rho(u,v)$ by the diameter. We can however normalise the correlation length with respect to the diameter. Calculating the mean over a variety of graph sizes gives \fref{figure9} which displays a divergent tendency at about a renormalised coupling value $\log(G_c)\approx 2$ for $D=2$ and $\log(G_c)\approx -1.4$ for $G=3$.
\begin{figure}[h]
	\centering
	\begin{subfigure}{0.45\textwidth}
		\includegraphics[width=\textwidth]{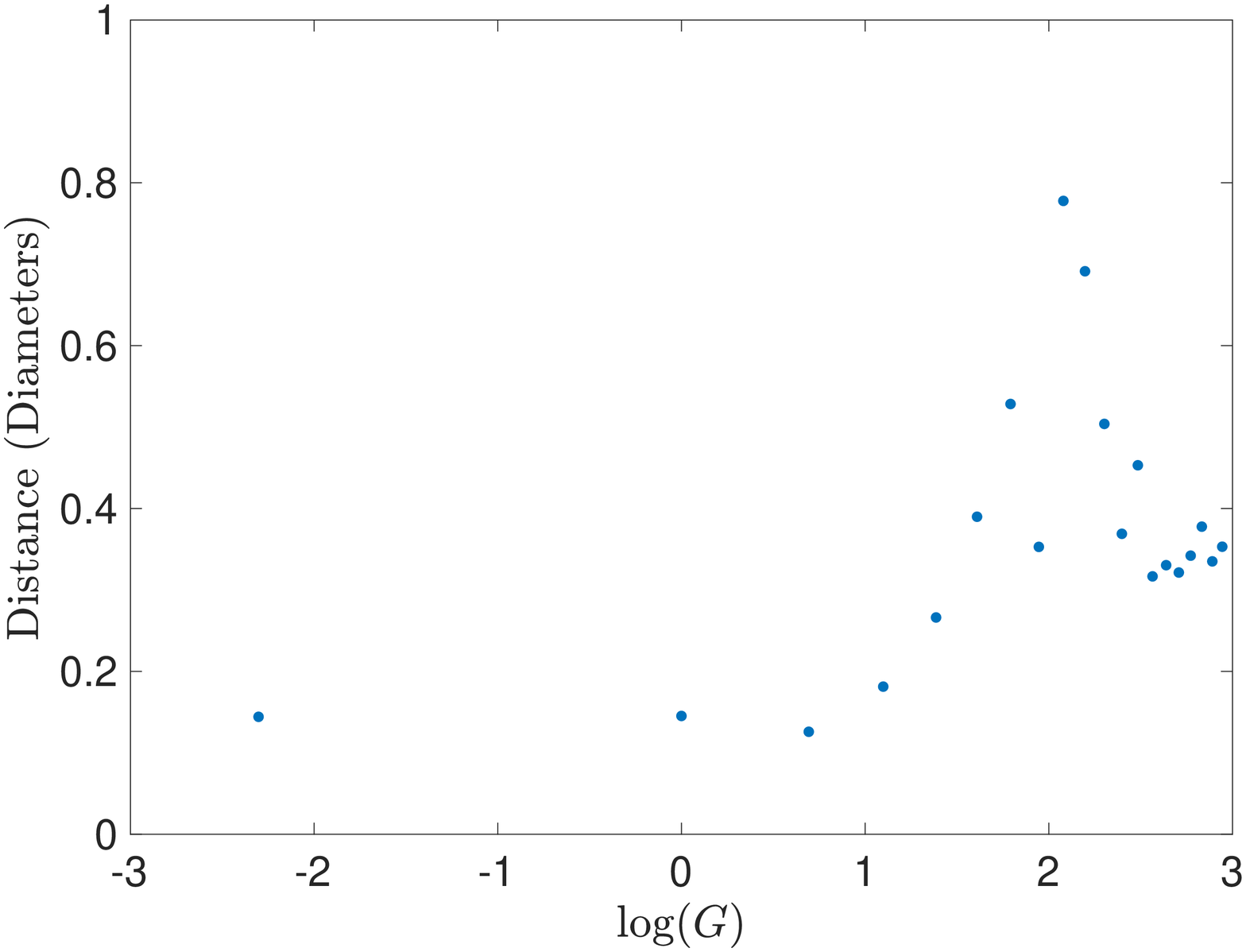}
		\caption{D=2.}\label{figure9a}
	\end{subfigure}
	\begin{subfigure}{0.45\textwidth}
		\includegraphics[width=\textwidth]{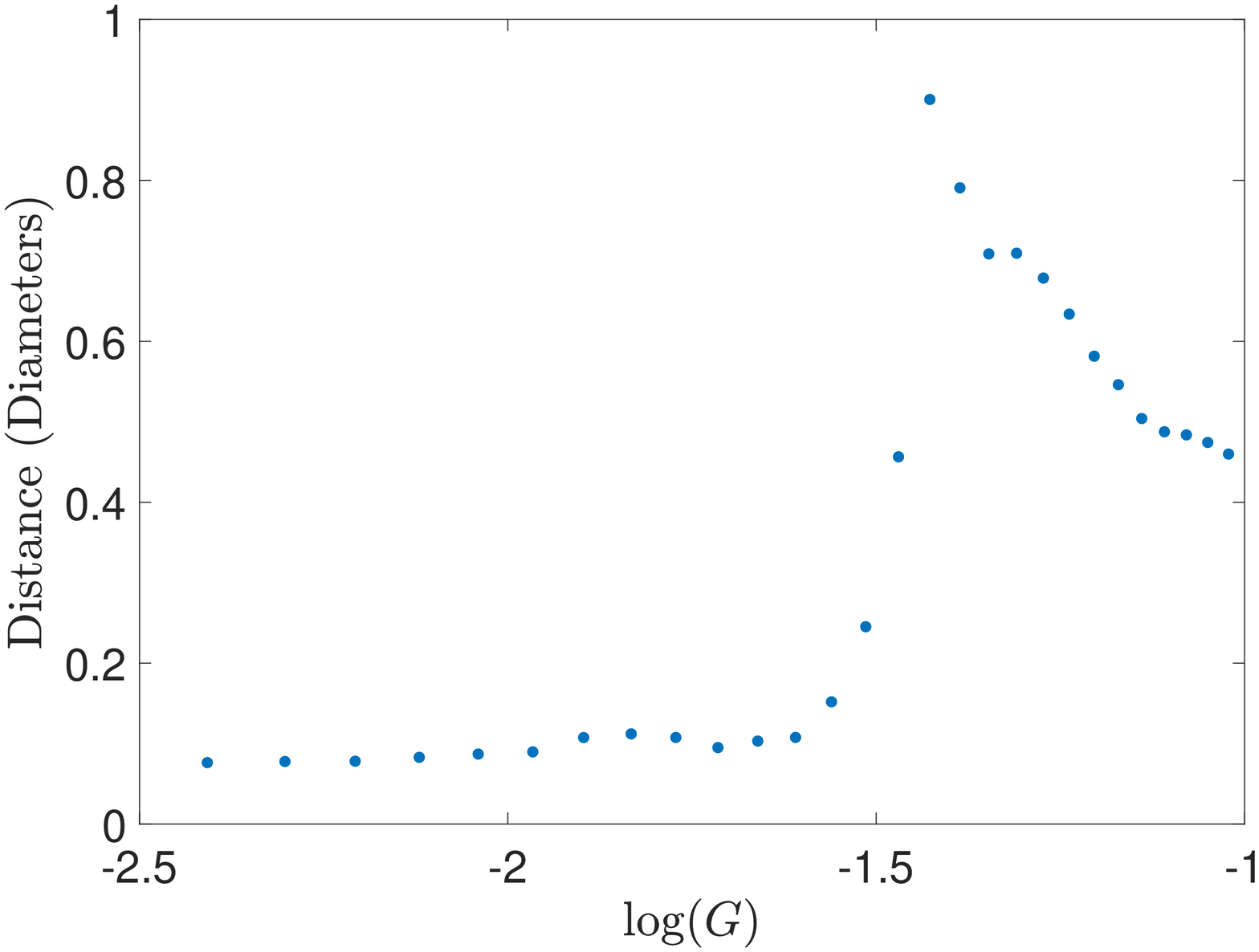}
		\caption{D=3.}\label{figure9b}
	\end{subfigure}
	\caption{Correlation length plots for $D=2$ and $D=3$ averaged over distinct graph sizes. There is reasonable indication of a divergence at $\log(G_c)\approx 2$ in $D=2$, and strong indication of a divergence at $\log(G_c)\approx -1.4$ in $D=3$, suggesting the existence of a continuous phase transition and consequently of a well-defined continuum limit in both these cases.}\label{figure9}
\end{figure}
\section{Conclusions}
Using methods from network theory and statistical mechanics, we have presented a discrete UV completion of Euclidean Einstein gravity, interpreted as a Euclidean quantum gravity theory. The Euclidean Einstein-Hilbert action is discretised via the \textit{Ollivier curvature} and to this end we have presented a new exact result for the Ollivier curvature in graphs; this should be of some interest to network theorists as the only exact expression that recognises the effect of more than one type of short cycle. Classical solutions of the action are Ollivier-Ricci flat configurations that approximate cubic complexes up to defects which are dynamically suppressed. Such configurations discretely approximate vacuum solutions to the Einstein field equations. It is shown that these configurations self-assemble from random graphs under the annealed Glauber dynamics associated with the action for low values of the quantum gravitational coupling $\hbar g$ and the Euclidean path integral is dominated by the `correct' phase in the classical limit. This improves on the situation in Euclidean dynamical triangulations, another attempt at Euclidean quantum gravity, insofar as classical solutions consisted of so-called baby universes. Baby universes do appear in our approach as solutions to \textit{mean field approximations} of our model, though such configurations are automatically suppressed when the analysis is exact. The dynamical suppression of defects and baby universe configurations are both automatic consequences of working in a configuration space defined by the hard core condition, which is thus sufficient for the stability of classical vacua. Finally, we find a continuous phase transition in a configuration space where geometric configurations are stable vacua, extending numerical results found in \cite{Trugenberger_CombQG}, and resolving another major stumbling block of the Euclidean dynamical triangulations programme. All these claims are supported by numerical evidence in $D=2$, while there is also evidence for the phase transition in $D=3$.

Despite the positive results presented herein, there remain several important conceptual points to be addressed. One central problem is the existence of non-trivial discrete manifolds and the generalisation of the above model to include cosmological constant terms or matter. (Another possibility is a generalisation in the other direction to include topology change by removing the regularity constraint; in \cite{Trugenberger_QGasNetSO}, one of the authors found regular graphs emerging from a rather natural Hamiltonian). As mentioned in the introduction, this procedure is closely related to weakening the hard core condition: by \eref{equation:NewOllivCurv}, only triangles can contribute positively to the curvature while it is argued in the above that the dynamical suppression of triangles is a consequence of hard core kinematics. The short independent cycle condition also appears somewhat artificial \textit{a priori}, at least from the gravitational side, and its main justification is utilitarian: given this assumption we find that exact calculations are possible and that spacetime graphs are stable. Nonetheless, the hard core condition does not seem to be essential for either of these results:  with regards to the mathematics, the core neighbourhood of an edge with independent short cycles is the simplest core neighbourhood to include triangles, squares and pentagons based on that edge. The method used to derive the main equation \eref{equation:NewOllivCurv}, however, easily generalises to less simple scenarios given a sufficiently refined partition of the core neighbourhood. In view of the complexity of the proof (see \ref{appendix:Proof}) in even this simple case, we have not attempted a more general derivation, but the authors believe in line with the discussion at the end of section \ref{subsection:OlivCurvII} that such a derivation is possible and suspect that the end result will not be qualitatively very different from the expression derived here. From a physical perspective, the hard core condition is essentially a \textit{dynamical }flat spacetime constraint and it is thus not surprising that this is \textit{sufficient} to stabilise space. Necessity is not as obvious, and at present it remains plausible that weaker conditions can play a similar stabilising role as the hard core condition. At the same time, the proper interpretation of a graph with non-negligible odd-cycle defects as a discrete manifolds appears rather subtle and the problem remains open. One possibility is the inclusion of configurations isomorphic to \fref{figure2a}: combinatorially we still have the local `isomorphism' with $\mathbb{Z}^D$ in a natural manner, but an optimal transport plan will `see' a triangle and thus such configurations contribute positively to the curvature. 

Another problem with the hard core condition is the computational difficulty of enforcing the constraint. Although it allows for significant improvements over the naive transport theoretic optimisation problem, the cost is an excluded subgraph constraint which also leads to severe limitations on the numerical analysis. Computational runtimes have prevented us from a more complete characterisation of the classical phase (via, for instance Hausdorff dimension calculations) or of the critical behaviour of the system. A finite size scaling analysis may well help to improve matters in this regard, though at present even such an analysis is precluded: an examination of \fref{figure8} at the coupling values around criticality (as obtained from \fref{figure9}) indicate that we are some way off the asymptotic regime.

The most significant shortcoming of the present model, however, is surely the \textit{a priori} Riemannian setting and the absence of any well-defined time dimension. One of the authors has elsewhere \cite{Trugenberger_RHLW} sketched a combinatorial argument whereby an emergent macroscopic dimension will obey a strong causality constraint. Difficulties in picking out such a dimension remain, however, and the problem of emergent time is far from resolved. Moreover, even with a successful resolution to this problem of time, the experience of Euclidean dynamical triangulations and other simplicial gravity approaches suggests that the interpretation of a Euclidean quantum gravity theory as a (covariantly) quantised model remains highly non-trivial: traditional flat spacetime problems such as reflection-positivity \cite{GlimmJaffe} are compounded by the fact that analytical continuation is not generally compatible with diffeomorphism invariance while it is also sometimes found that the analytically continued action is complex or has no lower bound and thus no stable vacua \cite{AGJL}. One of the triumphs of \textit{causal} dynamical triangulations over its Euclidean predecessor was the demonstration of rotatability of Lorentzian to Euclidean signature despite the insolubility of the above problems in the reverse direction. These are all issues which the present approach must ultimately address.
\ack 
C.K. acknowledges studentship funding from EPSRC under the grant number EP/L015110/1. F.B. acknowledges funding from EPSRC (UK) and the Max Planck Society for the Advancement of Science (Germany).

\appendix
\section{Some Notation and Terminology}\label{appendix:Notation}
We briefly recall graph theoretic terminology, essentially following \cite{Diestel_GraphTheory} though we occasionally depart from the conventions therein if the departure is relatively common usage. A \textit{simple graph} $\omega$ is a pair $(V(\omega),E(\omega))$ where $V(\omega)$ is a set of \textit{vertices} of $\omega$ and $E(\omega)\subseteq \mathcal{P}_2(V(\omega))$ is a set of \textit{edges} of $\omega$, where each edge is an unordered pair of vertices. That is to say $\mathcal{P}_2(X):=\set{x\subseteq X:|x|=2}$. A graph is said to be \textit{complete} iff $E(\omega)=\mathcal{P}_2(V(\omega))$. We denote $uv:=\set{u,v}$ for any $u,\:v\in V(\omega)$. $|V(\omega)|$ is the \textit{order} of $\omega$ and $|E(\omega)|$ is the \textit{size} of $\omega$. Two vertices $u,\:v\in V(\omega)$ are \textit{adjacent} or \textit{neighbours} iff $uv\in E(\omega)$; we denote the set of neighbours of $u$ by $N_\omega(u):=\set{v\in V(\omega):uv\in E(\omega)}$. The \textit{degree} of a vertex is the number of neighbours of that vertex; we write $d_\omega(u):=|N_\omega(u)|$ or when $\omega$ is clear we may simply write $d_u$. A graph is said to be \textit{$d$-regular} iff any two vertices have the degree $d$, while a graph is \textit{regular} iff it is $d$-regular for some $d\in \mathbb{N}$. A vertex $u\in V(\omega)$ and an edge $e\in E(\omega)$ are \textit{incident} iff $u\in e$ while two edges $e_1,\:e_2\in E(\omega)$ are incident iff $e_1\cap e_2\neq \emptyset$. A graph $\omega$ is said to be \textit{bipartite} iff there is a partition $(U,V)$ of $V(\omega)$ such that each edge $e\in E(\omega)$ is incident to exactly one edge in $U$ and one edge in $V$. $(U,V)$ is then called the \textit{bipartition} of $\omega$. A graph $\omega_1$ is a \textit{subgraph} of a graph $\omega_2$ iff $V(\omega_1)\subseteq V(\omega_2)$ and $E(\omega_1)\subseteq E(\omega_2)$; we write $\omega_1\subseteq \omega_2$. A subgraph is said to be \textit{induced} iff every edge of $\omega_2$ incident only to points in $\omega_1$ is an edge of $\omega_1$: $E(\omega_1)=E(\omega_2)\cap \mathcal{P}_2(V(\omega_1))$. 

A \textit{path} in a graph $\omega$ is a sequence $u_0u_1...u_n$ of vertices of $\omega$ such that $u_i\in N_\omega =(u_{i+1})$ for each $0\leq i< n$. A path $u_0...u_n$ is said to \textit{intersect} with a vertex $u$ iff $u=u_j$ for some $0\leq j\leq n$. The \textit{length} of the path $u_0...u_n$ is $n-1$ i.e. one less than the length of the sequence. We then say that $u_0...u_n$ is an $n$-path. A path is said to be \textit{simple} iff all vertices in the sequence are distinct. Note that elsewhere in the text, the term path implicitly means \textit{simple path} unless specified otherwise. A simple path $u_0...u_n$ is said to be a simple path \textit{between} the vertices $u_0$ and $u_n$. A simple path between $u,\:v\in V(\omega)$ is said to be \textit{geodesic} iff its length is less than or equal to the length of any other path between $u$ and $v$. The \textit{distance} between $u$ and $v$ is the length of any geodesic between $u$ and $v$, and is denoted $\rho_\omega(u,v)$. A graph is said to be \textit{connected} iff there is a finite path between any two vertices. The distance defines a metric function on $V(\omega)$ for any connected graph $\omega$. A \textit{circuit} is a path $u_0...u_n$ such that $u_0=u_n$. A circuit $u_0...u_{n}$, $u_0=u_n$, is said to be \textit{supported on an edge $uv$} iff $u=u_i$ and $v=u_{i+1}$ (or vice versa) for some $0\leq i<n$. Two circuits are said to \textit{share} an edge iff they are both supported on that edge. A circuit $u_0...u_n$ is a \textit{cycle} iff the $(n-1)$-path $u_0...u_{n-1}$ is simple. When there is no risk of confusion we may denote a cycle $u_0...u_n$ by the simple path $u_0...u_{n-1}$. The length of a cycle (more generally circuit) is simply its length as a path. We call $3$, $4$ and $5$-cycles \textit{triangles, squares} and \textit{pentagons} respectively. For any edge $uv\in E(\omega)$ we let $\triangle_{uv}$, $\square_{uv}$ and $\pentagon_{uv}$ denote the number of triangles, squares and pentagons supported on $uv$ respectively. A cycle is said to be \textit{odd} iff it has odd length. A graph is bipartite iff it has no odd cycles.

We introduce the following sets for any edge $uv\in E(\omega)$:
\begin{itemize}
	\item $N_\omega^v(u):=N_\omega(u)-\set{v}$ and symmetrically for $N_\omega^u(v)$.
	\item $\square(u,v)$ is the set of all vertices other than $u$ and $v$ intersecting a $4$-cycle supported on $uv$.
	\item $\pentagon(u,v)$ is the set of all vertices other than $u$ and $v$ intersecting a $5$-cycle supported on $uv$.
\end{itemize}
These lead to the following core neighbourhood partition sets:
\begin{enumerate}
	\item $\triangle(uv):=N_\omega(u)\cap N_\omega(v)$.
	\item $\square(u):=\square(u,v)\cap N_\omega^v(u)$ and similarly for $\square(v)$.
	\item $\pentagon(u):=\pentagon(u,v)\cap N_\omega^v(u)$ and similarly for $\pentagon(v)$.
	\item $\rm{Fr}(u):=N_\omega^v(u)-(\triangle(uv)\cup \square(u)\cup\pentagon(u))$ and similarly for $\rm{Fr}(v)$. We also define $n_u:=|\rm{Fr}(u)|$ and $n_v:=|\rm{Fr}(v)|$.
	\item $\pentagon(uv):=\set{w\in V(\omega):\rho_\omega(u,w)=2\rm{ and }\rho_\omega(v,w)=2}$.
\end{enumerate}
An edge $uv\in E(\omega)$ has independent short cycles iff any two short cycles on $uv$ share no other edges. For an edge with independent short cyclesm the core neighbourhood partition sets are pairwise disjoint and
\begin{eqnarray}
\triangle_{uv}=|\triangle(uv)| \qquad \square_{uv}=|\square(u)|=|\square(v)|\qquad  \pentagon_{uv}=|\pentagon(u)|=|\pentagon(v)|=|\pentagon(uv)|.
\end{eqnarray}
Also
\begin{eqnarray}\label{equation:Degree}
d_\omega(x)=1+\triangle_{uv}+\square_{uv}+\pentagon_{uv}+n_x
\end{eqnarray}
for any $x\in uv$.
\section{Proof of the Main Equation}\label{appendix:Proof}
By locality we consider the standard core neighbourhood $c_{uv}$ of an edge $uv\in E(\omega)$ with independent short cycles; transport profits and costs will be defined in this core neighbourhood unless otherwise specified and we drop the subscript $c_{uv}$ for convenience. We shall also drop subscript and superscript $uv$s since we deal only with the edge $uv$.

Recall that the distance matrix \eref{equation:DistanceMatrix} for the standard core neighbourhood of an edge with independent short cycles is given
\begin{eqnarray}
\begin{array}{c|ccccc}
\bi{D} & u & \triangle(uv) & \square(v) & \pentagon(v) & \rm{Fr}(v)\\
\hline
v & 1 & 1 & 1 & 1 & 1\\
\triangle(uv) & 1 & 0 & 2 & 2 & 2\\
\square(u) & 1 & 2 & 1 & 3 & 3\\
\pentagon(u) & 1 & 2 & 3 & 2 & 3\\
\rm{Fr}(u) & 1 & 2 & 3 & 3 & 3
\end{array}.\nonumber
\end{eqnarray}
Elements of any transport plan $\bpi$ take their values in the interval $[0,1]$ and so it is clear that the distance matrix contributes more significantly to the cost. Optimal transport plans will thus avoid transferring earth between distantly separated blocks. In particular any \textit{optimal} transport plan will move as much dirt as possible from $\rm{Fr}(u)$ and $v$ to $u$ and $\rm{Fr}(v)$ respectively. We thus consider distinct cases based on the ratios
\numparts
\begin{eqnarray}
R_u:=\frac{n_um_u}{m_v}\\
R_v:=\frac{n_vm_v}{m_u}
\end{eqnarray}
\endnumparts
where for convenience we define $m_u:=d_u^{-1}$ and $m_v:=d_v^{-1}$. Roughly speaking $R_u<1$ says that the total mass at $u$ can be used to fill $\rm{Fr}(u)$. Similarly $R_v<1$ implies that all the mass at $\rm{Fr}(v)$ can be contained at $u$. If $R_1,\:R_2> 1$, then mass from elsewhere must be transported to $\rm{Fr}(u)$ or the mass at $\rm{Fr}(v)$ must be transported elsewhere leading to an increased cost.

Naively, we have the following mutually exclusive cases:
\begin{eqnarray}
R_u,\:R_v<1\qquad R_u<1\leq R_v \qquad R_v<1\leq R_u \qquad 1\leq R_u,\:R_v.
\end{eqnarray}
However, not all these cases are consistent. In particular, first assume without loss of generality that
\begin{eqnarray}
d_u\geq d_v\qquad m_v\geq m_u.
\end{eqnarray}
Then since $1=d_um_u=d_vm_v$ by definition, we have
\begin{eqnarray}\label{equation:Inequality1}
(1+\triangle +\square +\pentagon +n_v)m_v=(1+\triangle +\square +\pentagon +n_u)m_u
\end{eqnarray}
by \eref{equation:Degree}, which we shall apply freely in the subsequent. Rearranging \eref{equation:Inequality1} gives
\begin{eqnarray}\label{inequality:MasterInequality}
\fl (\triangle+\square+\pentagon) (m_v-m_u)=(n_um_u-m_v)-(n_vm_v-m_u)\nonumber\\
=m_v(R_u-1)-m_u(R_v-1).
\end{eqnarray}
The LHS of the above equation is trivially non-negative, so $R_u<1$ implies $R_v<1$ and the analysis boils down to the following \textit{three} cases:
\begin{eqnarray}\label{Cases}
R_u,\:R_v<1\qquad R_v<1\leq R_u\qquad 1\leq R_u,\:R_v.
\end{eqnarray}
For each case we shall specify one or more transport plans (we may need to further subdivide cases) and demonstrate their optimality by showing that a bounded map with appropriate transport profit is $1$-Lipschitz.
\begin{enumerate}
	\item $R_u<1,\:R_v<1$.\\
	Consider the transport plan
	\begin{eqnarray}\label{equation:Pi1a}
	\fl \begin{array}{c||ccccc|c}
	{\bpi_1^a} & u & \triangle(uv) & \square(v) & \pentagon(v) & \rm{Fr}(v) & \\
	\hline\hline
	v & 0 &  \triangle(m_v-m_u) & \square(m_v-m_u) & \pentagon (m_v-m_u)+m_v-n_um_u & n_vm_v & m_u \\
	\triangle(uv) & 0 & \triangle m_u & 0 & 0 & 0 & \triangle m_u\\
	\square(u) & 0 & 0 & \square m_u & 0 & 0 & \square m_u\\
	\pentagon(u) & m_v-n_um_u & 0  & 0 & (\pentagon +n_u)m_u-m_v  & 0 & \pentagon m_u\\
	\rm{Fr}(u) & n_um_u & 0 & 0 & 0 & 0 & n_um_u\\
	\hline
	& m_v & \triangle m_v & \square m_v & \pentagon m_v & n_vm_v & 
	\end{array}.
	\end{eqnarray}
	We have explicitly included the row and column sums for this transport plan, though henceforth these shall be understood implicitly. It is immediately clear that the row and column sums are appropriate, except for the row headed by $v$. Here note that
	\begin{eqnarray}
	\fl \triangle(m_v-m_u)+\square (m_v-m_u)+(\pentagon (m_v-m_u)+m_v-n_um_u)+n_vm_v\\
	=(1+\triangle +\square +\pentagon+n_v)m_v -(\triangle +\square +\pentagon +n_u)m_u\nonumber\\
	=d_vm_v-(d_u-1 )m_u\nonumber\\
	=m_u\nonumber
	\end{eqnarray}
	as required. Note the condition $R_v<1$ is included implicitly in the specification of ${\bpi_1^a}$ since the top right entry assumes $n_vm_v<m_u$. The terms $n_um_u$, $\triangle m_u$, $\square m_u$ and $n_vm_v$ are trivially positive. The terms $\triangle (m_v-m_u)$ and $\square (m_v-m_u)$ are also positive due to the non-restrictive assumption $m_v\geq m_u$. The terms $m_v-n_um_u$ and $m_u-n_vm_v$ are positive because of the conditions $R_u,\:R_v<1$ which hold for the present case. $\pentagon (m_v-m_u)+(m_v-n_um_u)$ is positive since each of the bracketed terms is positive. Finally, $(\pentagon +n_u)m_u-m_v$ is not necessarily positive, and we must assume that it is for the current plan to exist:
	\begin{eqnarray}\label{Inequality1}
	0\leq (\pentagon +n_u)m_u-m_v\leq \pentagon m_v
	\end{eqnarray}
	where the RHS comes from the fact that we are able to transport $(\pentagon +n_u)m_u-m_v$ to $\pentagon (v)$. On the other hand it follows trivially since $(\pentagon +n_u)m_u-m_v=\pentagon m_u-(m_v-n_um_u)\leq \pentagon m_v-(m_v-n_um_u)\leq \pentagon m_v$ since $m_v\geq m_u$ and $R_v<1$. That is to say, ${\bpi_1^a}$ exists iff $R_u<1$, $R_v<1$ and the left hand inequality of expression \ref{Inequality1} holds.
	
	The transport cost associated with $\pi_1^a$ is:
	\begin{eqnarray}\label{equation:WPi1a}
	\fl W^{\pi_{1}^a}(u,v)=D_H(u,v)\cdot \pi_1^a(u,v)\nonumber\\
	=(-\triangle -\square -\pentagon -n_u+0\triangle +\square -n_u+2\pentagon +2 n_u+n_u)m_u+(\triangle +\square +\pentagon+1+n_v+1-2)m_v\nonumber\\
	=(-\triangle +\pentagon +n_u)m_u+(\triangle +\square +\pentagon +n_v)m_v.
	\end{eqnarray}
	We are looking for a short map $f_1^a$ such that $W^{f_1^a}(u,v)=W^{\pi_{1}^a}(u,v)$. Recalling that
	\begin{eqnarray}
	W^f(u,v)=m_u\sum_{w\in N(u)}f(w)-m_v\sum_{w\in N(v)}f(w)\nonumber,
	\end{eqnarray}
	we can read off the form of a map with appropriate transport profit from \eref{equation:WPi1a}. In particular, the mapping $f_1^a:V(c_{uv})\rightarrow \mathbb{Z}$ given by
	\begin{eqnarray}\label{equation:f1a}
	f_1^a(w)=\left\{\begin{array}{rl}
	1, & w\in \pentagon(u)\cup \rm{Fr}(u)\\
	0, & w\in uv\cup \square(u)\cup\pentagon(uv)\\
	-1, & w\in \triangle(uv)\cup \square(v)\cup\pentagon(v)\cup \rm{Fr}(v)
	\end{array}\right.
	\end{eqnarray}
	should have the `correct' transport profit. We may check this explicitly:
	\begin{eqnarray}\label{equation:Wf1a}
	\fl W^{f_1^a}(u,v)=m_u\sum_{w\in N(u)}{f_1^a}(w)-m_v\sum_{w\in N(v)}{f_1^a}(w)\nonumber\\
	=m_u\sum_{w\in \pentagon (u)\cup \rm{Fr}(u)}{f_1^a}(w)+m_u\sum_{w\in \set{v}\cup\square(u)}{f_1^a}(w)+m_u\sum_{w\in \triangle(uv)}f_1^a(w)\nonumber\\
	\qquad-m_v\sum_{w\in \set{u}}{f_1^a}(w) -m_v\sum_{w\in \triangle(uv)\cup\square(v)\cup \pentagon(v)\cup \rm{Fr}(v)}{f_1^a}(w)\nonumber\\
	=(\pentagon +n_u+0(1+\square)-\triangle)m_u-(0\cdot 1-\triangle -\square -\pentagon -n_v)m_v\nonumber\\
	=(-\triangle +\pentagon +n_u)m_u+(\triangle +\square +\pentagon+n_v)m_v.
	\end{eqnarray}
	Comparing equations \eref{equation:WPi1a} and \eref{equation:Wf1a} shows that
	\begin{eqnarray}\label{equation:W1a}
	\fl W^{1a}(u,v):=W^{\pi_1^a}(u,v)\nonumber\\
	=W^{f_1^a}(u,v)\nonumber\\
	=(-\triangle +\pentagon +n_u)m_u+(\triangle +\square +\pentagon+n_v)m_v\nonumber\\
	=(-\triangle +\pentagon +n_u)m_u+(d_v-1)m_v\nonumber\\
	=1-\triangle m_v+((\pentagon +n_v)m_v-m)
	\end{eqnarray}
	and we have found an expression for the Wasserstein metric in this scenario.
	
	Now consider the case where
	\begin{eqnarray}\label{Inequality2}
	(\pentagon +n_u)m_u-m_v< 0.
	\end{eqnarray}
	We examine the transferral plan
	\begin{eqnarray}\label{equation:Pi1b}
	\begin{array}{c||ccccc}
	{\bpi_1^b} & u & \triangle(uv) & \square(v) & \pentagon(v) & \rm{Fr}(v) \\
	\hline\hline
	v & m_v-(\pentagon +n_u)m_u &  \triangle(m_v-m_u) & \square(m_v-m_u) & \pentagon m_v & n_vm_v \\
	\triangle(uv) & 0 & \triangle m_u & 0 & 0 & 0 \\
	\square(u) & 0 & 0 & \square m_u & 0 & 0 \\
	\pentagon(u) & \pentagon m_u & 0  & 0 & 0 & 0 \\
	\rm{Fr}(u) & n_um_u & 0 & 0 & 0 & 0
	\end{array}.
	\end{eqnarray}
	It is immediately verified that this transport plan is well-defined and always exists for the case defined by the inequalities $R_u<1,\:R_v<1$ and inequality \ref{Inequality2}. ${\bpi_1^b}$ has a transport cost
	\begin{eqnarray}\label{equation:Cost_Pi^b_Ruv<1}
	\fl W^{\pi_1^b}(u,v)=(-\pentagon -n_u-\triangle -\square +0\triangle +\square +\pentagon +n_u)m_u+(1+\triangle +\square +\pentagon +n_v)m_v\nonumber\\
	=-\triangle m_u+(1+\triangle +\square +\pentagon +n_v)m_v.
	\end{eqnarray}
	We read off a map
	\begin{eqnarray}\label{equation:f^1b}
	f_1^b(w)=\left\{\begin{array}{rl}
	-1, &w\in  N(v)\\
	0, & w\in (N(u)-\triangle(u,v))\cup \pentagon(uv)
	\end{array}\right.
	\end{eqnarray}
	which has the appropriate transport profit, and immediately ascertain that it is short as required. Thus we have a Wasserstein distance
	\begin{eqnarray}\label{equation:W1b}
	\fl W^{1b}(u,v):=-\triangle m_u+(1+\triangle +\square +\pentagon +n_v)m_v\nonumber\\
	=1-\triangle m_u.
	\end{eqnarray}
	Comparing equations \eref{equation:W1a} and \eref{equation:W1b} in light of the defining inequalities immediately suggests a Wasserstein distance
	\begin{eqnarray}
	W^{1}(u,v):=1-\triangle m_u+[(\pentagon +n_u)m_u-m_v]_+.
	\end{eqnarray}
	where $[a]_+=\max(a,0)$; this is a single expression giving the Wasserstein distance whenever $R_{u},\:R_V<1$.
	\item $R_v<1\leq R_u$.\\
	The transport plan
	\begin{eqnarray}\label{equation:Pi^a_3}
	\begin{array}{c||ccccc}
	{\bpi^a_{2}} & u & \triangle(uv) & \square(v) & \pentagon(v) & \rm{Fr}(v) \\
	\hline\hline
	v & 0 & 0 & 0 & m_u-n_vm_v & n_vm_v \\
	\triangle(uv) & 0 & \triangle m_u & 0 & 0 & 0 \\
	\square(u) & 0 & 0 & \square m_u & 0 & 0 \\
	\pentagon(u) & 0 & 0  & 0 & \pentagon m_u & 0 \\
	\rm{Fr}(u) & m_v & \triangle (m_v-m_u) & \square (m_v-m_u) & (\pentagon +n_v)m_v-(1+\pentagon)m_u & 0
	\end{array}
	\end{eqnarray}
	exists whenever $(\pentagon +n_v)m_v\geq (1+\pentagon)m_u$. This has an associated transport cost
	\begin{eqnarray}
	\fl W^{\pi_2^a}(u,v)=(1+0\triangle +\square +2\pentagon -2\triangle -3\square -3\pentagon -3)m_u+(-n_v+n_v+m_v+2\triangle +3\square +3\pentagon +3n_v)m_v\nonumber\\
	=-(2+2\triangle +2\square +\pentagon )m_u+(1+2\triangle +3\square +3\pentagon +3n_v)m_v.
	\end{eqnarray}
	The map
	\begin{eqnarray}\label{equation:f2a}
	f_2^a(w)=\left\{\begin{array}{rl}
	0, & f\in \rm{Fr}(u)\\
	-1, & f\in \set{u}\cup \pentagon(u)\\
	-2, & f\in \set{v}\cup \triangle(u,v)\cup \square(u)\cup \pentagon(uv)\\
	-3, & f\in \square(v)\cup\pentagon(v)\cup \rm{Fr}(v)
	\end{array}\right.
	\end{eqnarray}
	satisfies $W^{f^a_2}(u,v)=W^{\pi_2^a}(u,v)$. It is simple to verify explicitly that $f^a_2$ is short; thus we may identify a Wasserstein distance
	\begin{eqnarray}
	\fl W^{2a}(u,v):=-(2+2\triangle +2\square +\pentagon )m_u+(1+2\triangle +3\square +3\pentagon +3n_v)m_v\nonumber\\
	=-(2d_u-\pentagon -2n_u)m_u+(3d_v-2-\triangle )m_v\nonumber\\
	=1-\triangle m_v+((\pentagon +n_u)m_u-m_v)+(n_um_u-m_v).
	\end{eqnarray}
	We now consider the case $(\pentagon +n_v)m_v<(1+\pentagon)m_u$, via the transport plan
	\begin{eqnarray}\label{equation:Pi^b_2}
	\fl \begin{array}{c||ccccc}
	{\bpi^b_{2}} & u & \triangle(uv) & \square(v) & \pentagon(v) & \rm{Fr}(v) \\
	\hline\hline
	v & 0 & 0 & (1+\pentagon)m_u-(\pentagon +n_v)m_v& \pentagon (m_v-m_u)  & n_vm_v\\
	\triangle(uv) & 0 & \triangle m_u & 0 & 0 & 0 \\
	\square(u) & 0 & 0 & \square m_u & 0 & 0\\
	\pentagon(u) & 0 & 0 & 0 & \pentagon m_u & 0 \\
	\rm{Fr}(u) & m_v & \triangle (m_v-m_u) & \square(m_v-m_u)-(1+\pentagon )m_u+(\pentagon +n_v)m_v & 0 & 0
	\end{array}
	\end{eqnarray}
	which exists as long as $\square (m_v-m_u)\geq (\pentagon +1)m_u-(\pentagon +n_v)m_v$. This has an associated cost
	\begin{eqnarray}
	\fl W^{\pi^b_2}(u,v)=(1+\pentagon-\pentagon +0\triangle +\square +2\pentagon -2\triangle -3\square -3-3\pentagon  )m_u\nonumber\\
	\qquad + (-\pentagon -n_v+\pentagon +n_v+1+2\triangle +3\square +3\pentagon +3 n_v)m_v\nonumber\\
	=-(2+2\triangle +2\square +\pentagon)m_u+(1+2\triangle +3\square +3\pentagon+n_v)m_v
	\end{eqnarray}
	We immediately see that $W^{\pi_2^b}(u,v)=W^{\pi_2^a}(u,v)$ and the transport plan is optimal by comparison with the transport cost of the $1$-Lipschitz map $f^a_2$ given in \eref{equation:f2a}.
	
	If $\square (m_v-m_u)< (\pentagon +1)m_u-(\pentagon +n_v)m_v$ but $\triangle(m_v-m_u)\geq (\pentagon +1)m_u-(\pentagon +n_v)m_v-\square (m_v-m_u)$ then we have the transport plan
	\begin{eqnarray}\label{equation:Pi^c_3}
	\begin{array}{c||ccccc}
	{\bpi^c_{2}} & u & \triangle(uv) & \square(v) & \pentagon(v) & \rm{Fr}(v) \\
	\hline\hline
	v & 0 & \alpha & \square(m_v-m_u) & \pentagon (m_v-m_u)  & n_vm_v \\
	\triangle(uv) & 0 & \triangle m_u & 0 & 0 & 0 \\
	\square(u) & 0 & 0 & \square m_u & 0 & 0 \\
	\pentagon(u) & 0 & 0 & 0 & \pentagon m_u & 0\\
	\rm{Fr}(u) & m_v & \beta  & 0 & 0 & 0 
	\end{array}
	\end{eqnarray}
	where we define
	\numparts
	\begin{eqnarray}
	\alpha:=(1+\pentagon )m_u-(\pentagon +n_v)m_v-\square(m_v-m_u)\\
	\beta:=\triangle (m_v-m_u) +\square(m_v-m_u)-(1+\pentagon )m_u+(\pentagon +n_v)m_v.
	\end{eqnarray}
	\endnumparts
	It has cost
	\begin{eqnarray}
	\fl W^{\pi_2^c}(u,v)=(1+\pentagon +\square -\square -\pentagon +0\triangle +\square +2\pentagon -2\triangle -2\square -2-2\pentagon )m_u\nonumber\\
	\qquad +(-\pentagon -n_v-\square +\square +\pentagon +n_v+1+2\triangle +2\square +2\pentagon +2n_v)m_v\nonumber\\
	=-(1+2\triangle +\square )m_u+(1+2\triangle +2\square +2\pentagon +2n_v)
	\end{eqnarray}
	which is the profit of the map
	\begin{eqnarray}
	f_2^c(w):=\left\{\begin{array}{rl}
	0, & \pentagon (u)\cup \rm{Fr}(u)\\
	-1, & uv\cup \square(u)\cup\pentagon(uv)\\
	-2, & \triangle(uv)\cup \square(v)\cup \pentagon (v)\cup \rm{Fr}(v).
	\end{array}\right.
	\end{eqnarray}
	This map is clearly short and so we have a Wasserstein distance
	\begin{eqnarray}
	\fl W^{2c}(u,v):=-(1+2\triangle +\square )m_u+(1+2\triangle +2\square +2\pentagon +2n_v)\nonumber\\
	=-(d_u+\triangle -\pentagon -n_u)m_u+(2d_v-1)m_v\nonumber\\
	=1-\triangle m_u+((\pentagon +n_u)m_u-m_v)
	\end{eqnarray}
	Comparing with $W^{\pi_1^a}(u,v)$ \eref{equation:WPi1a} given the short map $f_1^a$ \eref{equation:f1a} implies that the transport plan $\pi_2^c$ is optimal.
	
	Consider the condition $\triangle(m_v-m_u)\geq (\pentagon +1)m_u-(\pentagon +n_v)m_v-\square (m_v-m_u)$ for the existence of the transport plan ${\bpi_2^c}$; rearranging we obtain
	\begin{eqnarray}
	(\triangle +\square +\pentagon +n_v)m_v\geq (1+\triangle +\square +\pentagon)m_u\nonumber
	\end{eqnarray}
	which implies that $1-m_v\geq 1-n_um_u$ i.e. $n_um_u\geq m_v$ or equivalently $R_u\geq 1$. Since this holds by hypothesis, we have two expressions $W^{2a}(u,v)$ and $W^{2c}(u,v)$ that hold for the case $R_v<1\leq R_u$ corresponding to situations where $\square (m_v-m_u)\geq (\pentagon +1)m_u-(\pentagon +n_v)m_v$ and $\square (m_v-m_u)< (\pentagon +1)m_u-(\pentagon +n_v)m_v$ respectively. Rewriting these inequalities we obtain
	\numparts
	\begin{eqnarray}
	(n_um_u-m_v)\geq \triangle (m_v-m_u) \\
	(n_um_u-m_v)< \triangle (m_v-m_u).
	\end{eqnarray}
	\endnumparts
	We can then summarise both scenarios with the expression:
	\begin{eqnarray}
	\fl W^2(u,v):=1-\triangle m_v+((\pentagon +n_u)m_u-m_v)+\triangle (m_v-m_u)\lor(n_um_u-m_v).
	\end{eqnarray}
	\item $R_v\geq 1$, $R_u\geq 1$.\\
	\begin{eqnarray}\label{equation:Pi_4}
	\begin{array}{c||ccccc}
	{\bpi_{3}} & u & \triangle(uv) & \square(v) & \pentagon(v) & \rm{Fr}(v) \\
	\hline\hline
	v & 0 & 0 & 0 & 0 & m_u\\
	\triangle(uv) & 0 & \triangle m_u & 0 & 0 & 0\\
	\square(u) & 0 & 0 & \square m_u & 0 & 0\\
	\pentagon(u) & 0 & 0  & 0 & \pentagon m_u & 0 \\
	\rm{Fr}(u) & m_v & \triangle (m_v-m_u) & \square (m_v-m_u) & \pentagon(m_v-m_u) & n_vm_v-m_u 
	\end{array}.
	\end{eqnarray}
	This plan is always well defined. The transport cost is:
	\begin{eqnarray}
	\fl W_H^{\pi_3}(u,v)=(1+0\triangle +\square +2\pentagon -2\triangle -3\square -3\pentagon-3)m_u +(1+2\triangle +3\square +3\pentagon +3n_v)m_v\nonumber\\
	=-(2+2\triangle +2\square +\pentagon )m_u+(1+2\triangle +3\square +3\pentagon +3n_v)m_v.
	\end{eqnarray}
	The fact that $W^{\pi_3}(u,v)=W^{\pi_2^a}(u,v)$ allows us to compare with the 1-Lipschitz map $f^a_2$ in \eref{equation:f2a} and implies a Wasserstein distance:
	\begin{eqnarray}
	\fl W^3(u,v):=-(2+2\triangle +2\square +\pentagon )m_u+(1+2\triangle +3\square +3\pentagon +3n_v)m_v\nonumber\\
	=1-\triangle m_v+((\pentagon +n_u)m_u-m_v)+(n_um_u-m_v)
	\end{eqnarray}
\end{enumerate}
We now seek a single expression $W_H(u,v)$ for the $3$ cases defined in \ref{Cases}. Recall:
\numparts
\begin{eqnarray}
R_u,\:R_v<1\Rightarrow W(u,v)=1+[(\pentagon +n_u)-m_v]_+-\triangle m_u\nonumber\\
R_v<1\leq R_u\Rightarrow W(u,v)=1+((\pentagon +n_u)m_u-m_v)-\triangle m_v+\triangle (m_v-m_u)\lor(n_um_u-m_v).\nonumber\\
1\leq R_u,\:R_v \Rightarrow W(u,v)=1+((\pentagon +n_u)m_u-m_v)-\triangle m_v+(n_um_u-m_v)\nonumber.
\end{eqnarray}
\endnumparts
The first thing to note is that for cases (ii) and (iii), $R_u\geq 1$ and the terms $(n_um_u-m_v)$ and $((\pentagon +n_u)m_u-m_v)$ are trivially non-negative, while the former vanishes and the latter is uncertain for $R_u<1$. Thus we may express
\begin{eqnarray}
W_H(u,v)=1+[(\pentagon +n_u)-m_v]_++\rm{extra\: terms}\nonumber
\end{eqnarray}
in each of the three cases. Since case (ii) splits naturally between case (i) and case (iii) given this form, it seems plausible that the extra terms are simply those of case (ii). This follows if $R_u,\:R_v<1$ implies that $\triangle (m_v-m_u)> (n_um_u-m_v)$ and $1\leq R_u,\:R_v$ implies $(n_um_u-m_v)\geq \triangle (m_v-m_u)$. The first claim---$R_u,\:R_v<1$ implies that $\triangle (m_v-m_u)> (n_um_u-m_v)$---follows immediately from the fact that $\triangle (m_v-m_u)$ is positive and $n_um_u-m_v$ is negative given $R_u<1$. For the second claim---$1\leq R_u,\:R_v$ implies $(n_um_u-m_v)\geq \triangle (m_v-m_u)$---simply note that
\begin{eqnarray}
 \triangle (m_v-m_u)\leq (\triangle+\square +\pentagon)(m_v-m_u)\leq (n_um_u-m_v)-(n_vm_v-m_u)\leq n_um_u-m_v\nonumber
\end{eqnarray}
where the central inequality holds since $d_u\geq d_v$ and the final inequality holds because $(n_vm_v-m_u)\geq 0$ iff $R_v\geq 1$. Thus we may define
\begin{eqnarray}
 W_H(u,v):=1-\triangle m_v+[(\pentagon +n_u)m_u-m_v]_++\triangle (m_v-m_u)\lor(n_um_u-m_v).
\end{eqnarray}
Finally we loosen the requirement that $d_u\geq d_v$ and write in a symmetric form i.e. we make the replacements $d_u\mapsto d_u\lor d_v$ and $d_v\mapsto d_u\land d_v$:
\begin{eqnarray}
\fl W(u,v):=1-\frac{\triangle} {d_u\land d_v}+\left[\frac{\pentagon +n_u\lor n_v}{d_u\lor d_v}-\frac{1}{ d_u\land d_v}\right]_++\left[\triangle \left(\frac{1}{d_u\land d_v}-\frac{1}{d_u\lor d_v}\right)\right]\lor \left(\frac{n_u\lor n_v}{d_u\lor d_v}-\frac{1}{d_u\land d_v}\right).
\end{eqnarray}
Equation \eref{equation:NewOllivCurv} then follows from the fact that $\kappa(uv)=1-W(u,v)$ and elementary rearrangements in light of \eref{equation:Degree}.

\bibliographystyle{unsrt}
\bibliography{Ref}

\end{document}